\documentclass[preprint,12pt,authoryear]{elsarticle}

\usepackage{amssymb}

\usepackage{multirow}
\usepackage{multicol}
\usepackage{booktabs}
\usepackage{makecell}
\usepackage{hyperref}

\usepackage{subfig}
\usepackage{hhline}
\usepackage{xkeyval}
\makeatletter
\newlength{\sfp@hseplen}\newlength{\sfp@vseplen}
\define@cmdkey{subfigpos}[sfp@]{pos}[ul]{}
\define@cmdkey{subfigpos}[sfp@]{font}[\small]{}
\define@cmdkey{subfigpos}[sfp@]{vsep}[5pt]{\setlength{\sfp@vseplen}{\sfp@vsep}}
\define@cmdkey{subfigpos}[sfp@]{hsep}[5pt]{\setlength{\sfp@hseplen}{\sfp@hsep}}
\newcommand{\subfigimg}[3][,]{%
  \setkeys{Gin,subfigpos}{pos,font,vsep,hsep,#1}
  \setbox1=\hbox{\includegraphics{#3}}
  \ifnum\pdfstrcmp{\sfp@pos}{ul}=0
    \leavevmode\rlap{\usebox1}
    \rlap{\hspace*{\sfp@hsep}\raisebox{\dimexpr\ht1-\sfp@vsep}{\sfp@font{#2}}}
    \phantom{\usebox1}
  \else\ifnum\pdfstrcmp{\sfp@pos}{ur}=0
    \leavevmode\usebox1
    \llap{\raisebox{\dimexpr\ht1-\sfp@vsep}{\sfp@font{#2}}\hspace*{\sfp@hsep}}
  \else\ifnum\pdfstrcmp{\sfp@pos}{lr}=0
    \leavevmode\usebox1
    \llap{\raisebox{\sfp@vsep}{\sfp@font{#2}}\hspace*{\sfp@hsep}}
  \else
    \leavevmode\rlap{\usebox1}
    \rlap{\hspace*{\sfp@hseplen}\raisebox{\sfp@vsep}{\sfp@font{#2}}}
    \phantom{\usebox1}
  \fi\fi\fi
}
\makeatother

\journal{Computerized Medical Imaging and Graphics}

\begin{document}

\noindent This article has been accepted for publication in \emph{Computerized Medical Imaging and Graphics}.\\
DOI: https://doi.org/10.1016/j.compmedimag.2023.102259\\
\newpage

\begin{frontmatter}



\title{Improving Segmentation and Detection of\\Lesions in CT Scans\\Using Intensity Distribution Supervision}


\author[inst1]{Seung Yeon Shin\corref{cor1}}
\cortext[cor1]{Corresponding author: ssy2280@gmail.com}
\author[inst1]{Thomas C. Shen}
\author[inst1]{Ronald M. Summers}

\affiliation[inst1]{organization={Imaging Biomarkers and Computer-Aided Diagnosis Laboratory, Radiology and Imaging Sciences},
            addressline={Clinical Center, National Institutes of Health}, 
            city={Bethesda},
            postcode={20892}, 
            state={MD},
            country={USA}}
            
\begin{abstract}
We propose a method to incorporate the intensity information of a target lesion on CT scans in training segmentation and detection networks.
We first build an intensity-based lesion probability (ILP) function from an intensity histogram of the target lesion.
It is used to compute the probability of being the lesion for each voxel based on its intensity.
Finally, the computed ILP map of each input CT scan is provided as additional supervision for network training, which aims to inform the network about possible lesion locations in terms of intensity values at no additional labeling cost.
The method was applied to improve the segmentation of three different lesion types, namely, small bowel carcinoid tumor, kidney tumor, and lung nodule. The effectiveness of the proposed method on a detection task was also investigated. We observed improvements of 41.3\% $\rightarrow$ 47.8\%, 74.2\% $\rightarrow$ 76.0\%, and 26.4\% $\rightarrow$ 32.7\% in segmenting small bowel carcinoid tumor, kidney tumor, and lung nodule, respectively, in terms of per case Dice scores. An improvement of 64.6\% $\rightarrow$ 75.5\% was achieved in detecting kidney tumors in terms of average precision.
The results of different usages of the ILP map and the effect of varied amount of training data are also presented.
\end{abstract}

\begin{graphicalabstract}
\centering
\includegraphics[width = 1\columnwidth]{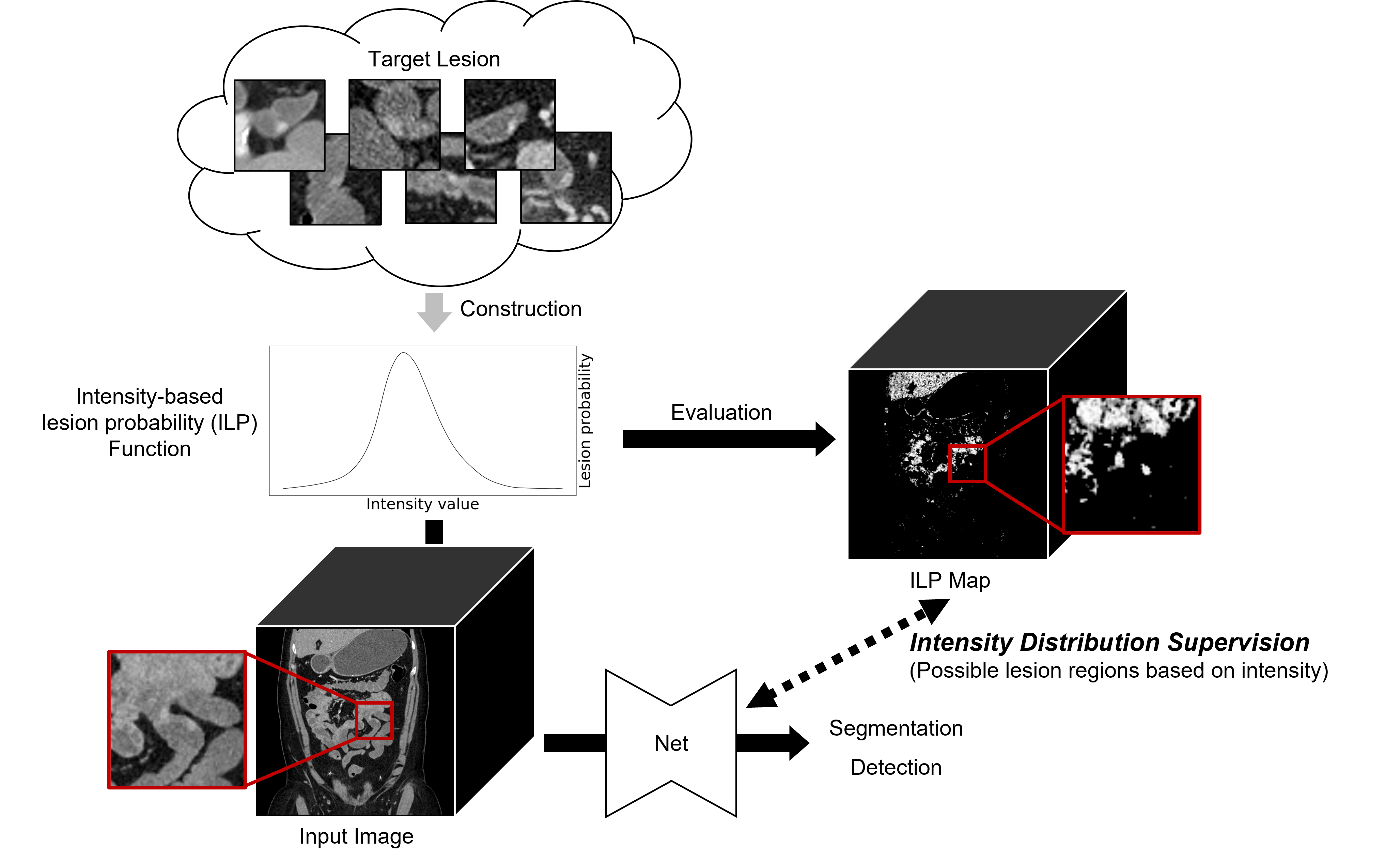}
\end{graphicalabstract}

\begin{highlights}
\item Intensity values in CT scans, i.e., Hounsfield unit, convey important information.
\item A method to incorporate the intensity information of a target lesion in training is proposed.
\item An intensity distribution of a target lesion is used to define an auxiliary task.
\item It informs the network about possible lesion locations based on intensity values.
\item A relative improvement of 2.4\% (16.9\%) was obtained in segmenting (detecting) kidney tumors.
\end{highlights}


\end{frontmatter}




\section{Introduction}\label{sec:intro}
Identification and quantification of abnormalities are the first objectives of medical image acquisition~\citep{eisenhauer09,shin19,ryu21}. According to the result, immediate treatments can be done, or follow-up studies are triggered for surveillance~\citep{cai21}. 

Abnormalities can arise at various locations in the body, such as different tissues and organs. When finding a particular type of lesion, it is often coupled with segmentation of an organ/tissue that can contain the lesion, e.g., liver and liver tumor segmentation~\citep{ayalew21}. These two related tasks can be considered either sequentially or jointly. In the first case, organ segmentation can benefit the following lesion identification by restricting the region of interest and thus by enabling more detailed inspection within it~\citep{shin15,kamble20}. Meanwhile, the latter facilitates joint optimization of the relevant tasks and thus can lead to an enhancement of them. In the work of \citet{tang20}, liver and liver tumor segmentations are performed jointly using a single shared network to utilize the correlation between them. Features learned for organ segmentation would be relevant to contained lesions since they reside within the organ.

Nevertheless, it is not always possible to train the organ segmentation together with the target lesion segmentation since it requires additional ground-truth (GT) segmentations of the organ. Indeed, a clinician aims to find and mark abnormalities, but not an entire organ, which hinders the mentioned joint or sequential modeling in most cases.

To boost the lesion segmentation without requiring additional supervision, there have been many attempts~\citep{weninger20,liu21,ma20}. \citet{weninger20} utilized a multi-task learning network, which performs an image reconstruction task in parallel with tumor segmentation, for brain tumors in magnetic resonance imaging (MRI) scans. By having a shared encoder and separate decoders for each task, training a part of the network, i.e., auto-encoder part, using scans without annotations is enabled. In the work of \citet{liu21}, lesion edge prediction is used as an auxiliary task to help the segmentation of skin lesions. Two tasks can interact with each other within the network and boost each other’s performance in turn. As a noticeable trend, incorporating distance transform maps of GT into segmentation networks has been tried in many different works in past years, which were summarized in the work of \citet{ma20}. Interestingly, most of the benchmark methods showed no or minor improvement against the baseline for tumor segmentation while they performed more preferably for organ segmentation. Challenges of tumor segmentation compared to organ segmentation, such as various locations, shapes, and sizes, are mentioned as a potential reason.

In terms of lesion detection, \citet{jaeger20} developed the Retina U-Net architecture based on the Retina Net~\citep{lin17_iccv}. The decoding part of the Retina Net is augmented by additional high-resolution feature levels, and semantic lesion segmentation is performed on top of them to boost the detection task. Despite its effectiveness, it assumes that segmentation annotations are available together with detection annotations, which is not always the case.

In computed tomography (CT) scans, intensity values, i.e., Hounsfield unit (HU), convey important information on the substance of each region, e.g., air, fat, and bone~\citep{buzug11}. Therefore, they can be used in identifying a particular organ, tissue, or lesion~\citep{petersenn15,lin16,phan18,summers06}. In the work of \citet{petersenn15}, a HU threshold of 13 or 21 is suggested to discriminate malignant adrenal tumors from benign ones in unenhanced CT scans. In the work of \citet{lin16}, a combined use of lesion morphology and HU values improved the
diagnostic accuracy in differentiating benign and malignant incidental breast lesions on contrast-enhanced chest CT scans. In the work of \citet{phan18}, a specific threshold range of $[40, 90]$ is used to determine areas of hemorrhage on brain CT scans.

In this paper, we propose a method to incorporate the intensity information of a target lesion on CT scans in training segmentation and detection networks. Instead of using hard thresholds as in the previous works~\citep{petersenn15,lin16,phan18}, an intensity distribution of a target lesion is first built and used to effectively locate regions where the lesions are possibly situated. The intensity distribution can be achieved by investigating intensity values within available GT lesion segmentations or can be provided as prior information.
More specifically, an intensity-based lesion probability (ILP) function constructed from an intensity histogram is used to compute the probability of being lesion for each voxel, and this soft label map is provided for network training as an auxiliary task.
It informs the network about our region of interest, which could contain target lesions, based on the intensity. Compared to the organ segmentation trained jointly with lesion identification tasks, our new task can be understood as a soft and possibly disconnected surrogate of organ segmentation, and it requires no additional annotation cost. 

We demonstrate the effectiveness of the proposed method by conducting experiments on three different datasets: 1) an in-house small bowel carcinoid tumor dataset, 2) the KiTS21 dataset~\citep{heller21} for kidney tumors, and 3) the LNDb dataset for lung nodules~\citep{pedrosa19}. The main contributions of our work are as follows. (1) We extend the idea of our previous paper~\citep{shin23_spie} to the segmentation of different lesions at different body locations to verify its generability. (2) We further investigate the effectiveness of the proposed method in several aspects, namely with varied amount of training data, in comparison to the joint organ segmentation, and even on a detection task.

\section{Datasets}\label{sec:datasets}
\subsection{Small Bowel Carcinoid Tumor Dataset}
Carcinoid tumor is a rare neoplasm (small bowel neoplasms including carcinoid tumors account for $0.5\%$ of all cancers in the United States~\citep{jasti20}) and found predominantly within the gastrointestinal tract ($50-71.4\%$) and especially in the small bowel ($24-44\%$)~\citep{hughes16}. They are often less than a centimeter in size~\citep{hughes16}.

Our carcinoid tumor dataset is composed of 24 preoperative abdominal CT scans collected at the National Institutes of Health Clinical Center. Each scan is from a unique patient who underwent surgery and had at least one carcinoid tumor within the small bowel. We note that creating a large dataset for small bowel carcinoid tumors is more difficult than for other more prevalent diseases.

All scans are intravenous and oral contrast-enhanced. An oral contrast agent of Volumen was used. Each patient has both arterial and venous phase scans, and either of them was selectively used according to the relevant description in the corresponding radiology report (18 arterial and 6 venous phase scans). They were acquired using 0.5, 1, or 2 mm slice thickness. All scans were cropped manually along the z-axis to include from the diaphragm through the pelvis. We will call this the SBCT dataset.

To achieve GT segmentation of tumors, we used ``Segment Editor" module in 3DSlicer~\citep{fedorov12}. The corresponding radiology report and an available $^{18}$F-DOPA PET scan were referred to for help in locating tumors. 88 tumors were annotated in total. We use five-fold cross-validation for this dataset.

\subsection{The KiTS21 Dataset}
Kidney cancer is the sixth and the ninth most common cancer for men and women, respectively, in the United States~\citep{cancernet_kidney_cancer}. 
The KiTS21 dataset aims to accelerate the development of automatic segmentation tools for renal tumors and surrounding anatomy. The KiTS21 cohort includes patients who underwent nephrectomy for suspected renal malignancy~\citep{heller21}. Preoperative CT scans of these patients were collected to compose the dataset. The official training set comprises 300 CT scans from 300 unique patients who had at least one kidney tumor. 

All scans are contrast-enhanced and were acquired in the late arterial phase. Every scan has corresponding GT segmentations of the kidney, tumor, and cyst. In this work, we focus on segmenting tumors while leaving cysts unattended since cysts are benign and clinically less relevant than tumors. We refer the authors to the dataset description paper~\citep{heller21} for more information. For experiments, we divide the dataset into training/validation/test sets at a ratio of 7:1:2.

\subsection{The LNDb Dataset}
Lung cancer is the leading cause of cancer death, which makes up almost $25\%$ of all cancer deaths~\citep{acs_lung_cancer}. Being a possible indicator of lung cancer, lung nodules show various shapes and characteristics. Thus, the identification and characterization of them are not trivial and prone to high inter-observer variability~\citep{pedrosa19}.

The LNDb dataset includes 294 intravenous contrast-enhanced CT scans from 294 unique patients. Fifty-eight scans among them were withheld by the organizers for the test set and the remaining 236 scans are available. Among all identified lesions (nodule $\geq$ 3 mm, nodule $<$ 3 mm, non-nodule), only nodules that are greater than or equal to 3 mm were segmented during the annotation process, and they will be segmented in this work. Thirty-five scans have been excluded since they have an empty segmentation map with the above-mentioned reason, resulting in 201 remained scans ($236 - 35 = 201$). 

In this work, we especially focus on improving the segmentation of \emph{non-solid} nodules since they are more likely to be malignant than \emph{solid} nodules and more difficult to identify due to their fuzzy appearance and lower incidence~\citep{diederich09}. We utilized nodule texture ratings ($1 - 5$) provided in the dataset, where 1 denotes closer to \emph{non-solid} nodules and 5 denotes closer to \emph{solid} nodules. According to these ratings, we classified each segmented nodule into two groups, namely, \emph{non-solid} ($\leq 2$) or \emph{solid} ($> 2$). Nineteen scans were identified to have at least one \emph{non-solid} nodule after manual inspection. We note that these 19 scans can also contain \emph{solid} nodules. Finally, they are used for two-fold cross-validation while the remaining 182 ($= 201 - 19$) scans are included as training images for every fold training.

\section{Methods}\label{sec:methods}
\subsection{Intensity Distribution Supervision}\label{subsec:int_distr}
\begin{figure}[t]
	\begin{center}
    \begin{minipage}{1\textwidth}
        \subfloat[]{\includegraphics[width = 0.5\textwidth]{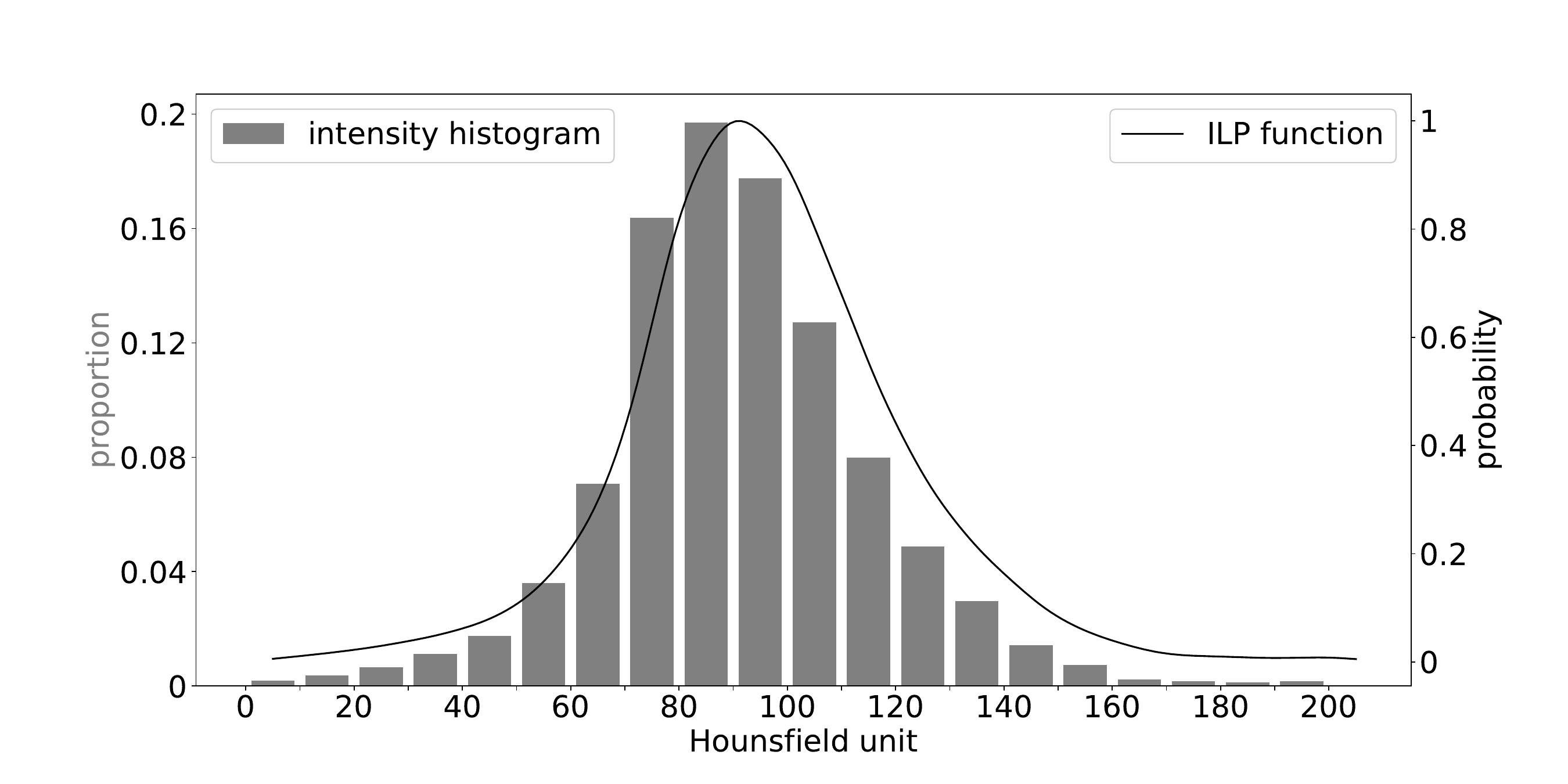}}
	    \subfloat[]{\includegraphics[width = 0.5\textwidth]{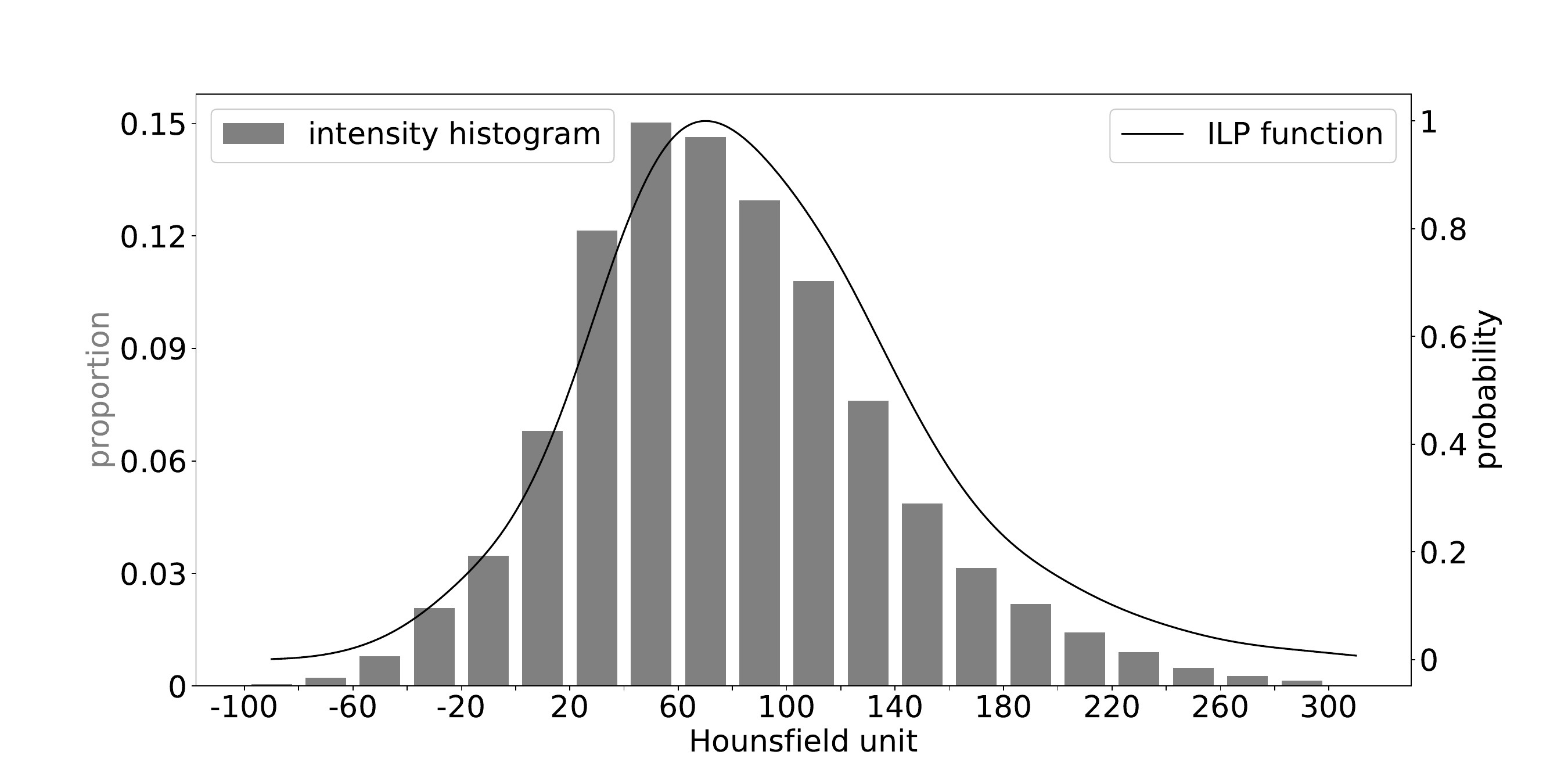}}
        \vspace{-0.4cm}
    \end{minipage}
    \begin{minipage}{1\textwidth}
	    \subfloat[]{\includegraphics[width = 0.5\textwidth]{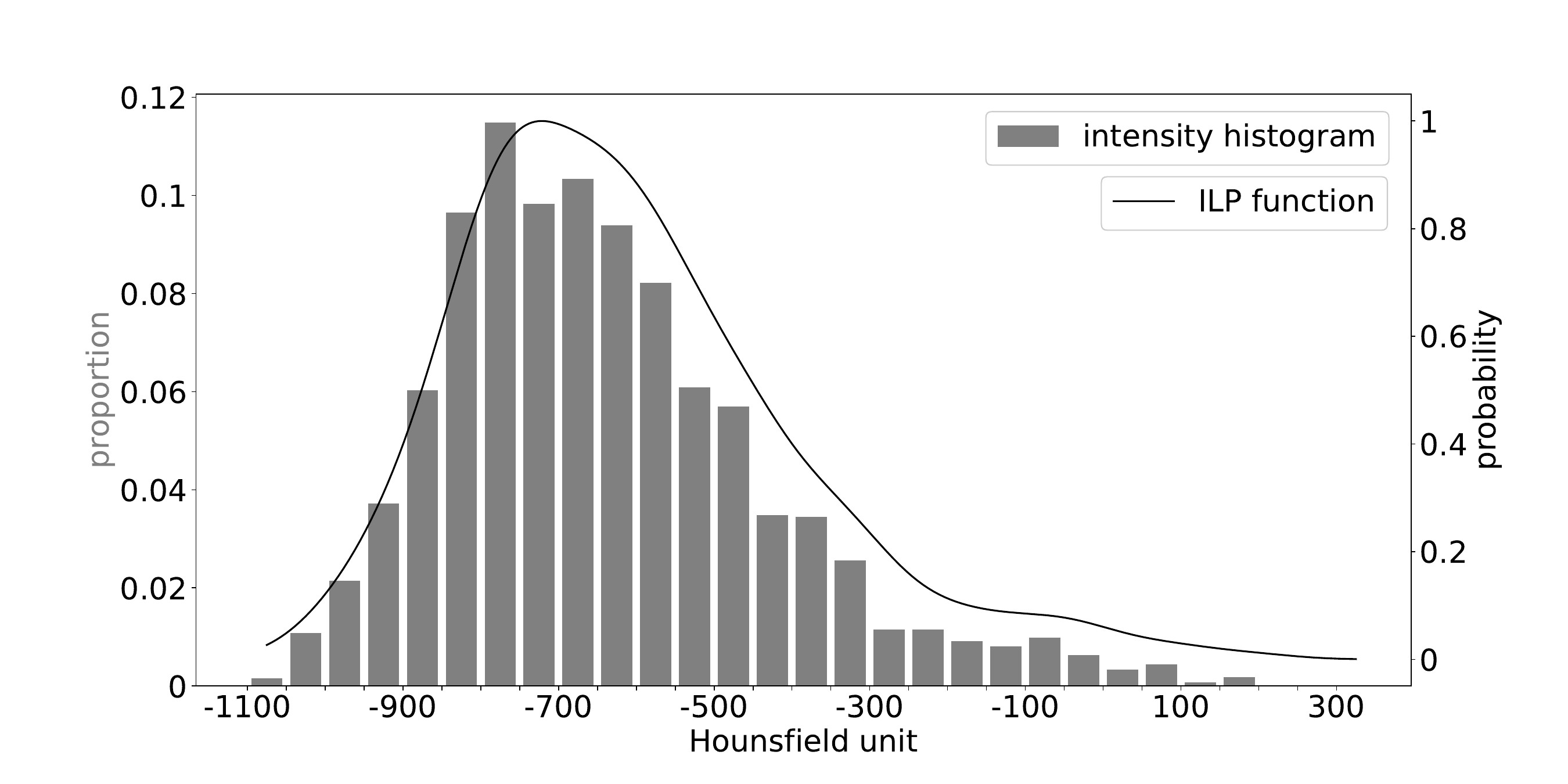}}
	    \subfloat[]{\includegraphics[width = 0.5\textwidth]{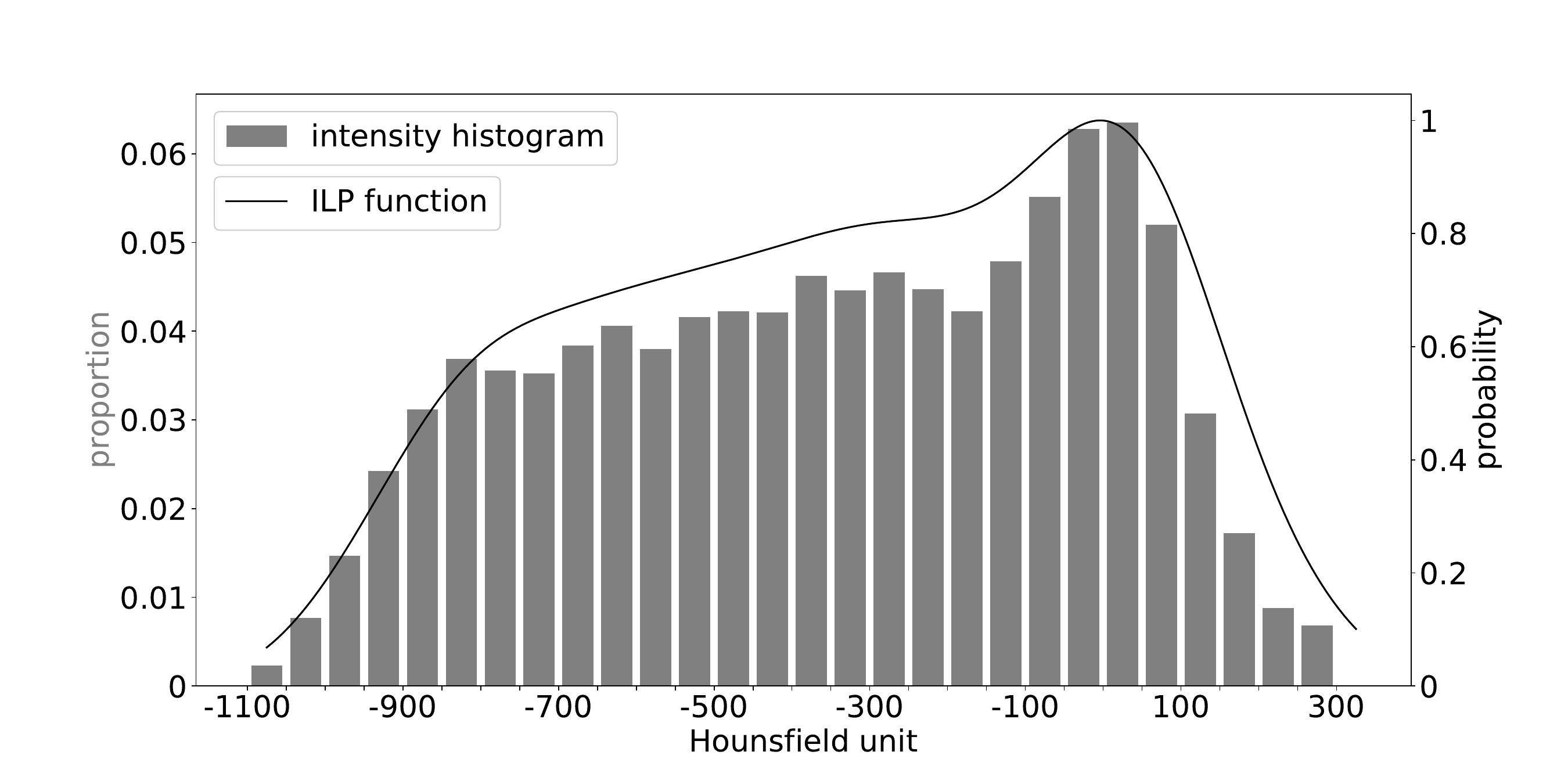}}
    \end{minipage}
	\caption{
    Intensity histogram (gray) and an intensity-based lesion probability (ILP) function (black) of target lesions for each dataset. 
    (a) Small bowel carcinoid tumors of the SBCT dataset. (b) Kidney tumors of the KiTS21 dataset. (c) \emph{Non-solid} lung nodules of the LNDb dataset. (d) \emph{Solid} lung nodules of the LNDb dataset. This is provided for comparison to that of \emph{non-solid} nodules.
    In each sub-figure, the intensity histogram and the ILP function have different scales, so the left or right y-axis should be read for each. Refer to the text for an explanation of the ILP function.
    }
    \label{fig:ilp_func} 
	\end{center}
\end{figure}

Figure~\ref{fig:ilp_func} presents the intensity histogram of target lesions of each dataset. They were computed by aggregating intensity values within GT lesion segmentations of each dataset. Images were smoothed using anisotropic diffusion~\citep{black98} before the histogram construction.
To make a smooth evaluable function from the discontinuous histogram, we perform kernel density estimation~\citep{parzen62}. It is a method used to estimate the probability density function based on kernels as basis functions. We used `gaussian\_kde' function of SciPy Python library, which uses Gaussian kernels with automatic bandwidth determination.
The resulting function is then rescaled to have the maximum value of $1$.
While it could be less precise, the ILP function can be provided also by a user as prior information.

\begin{figure}[t]
	\centering
	\includegraphics[width = 1\columnwidth]{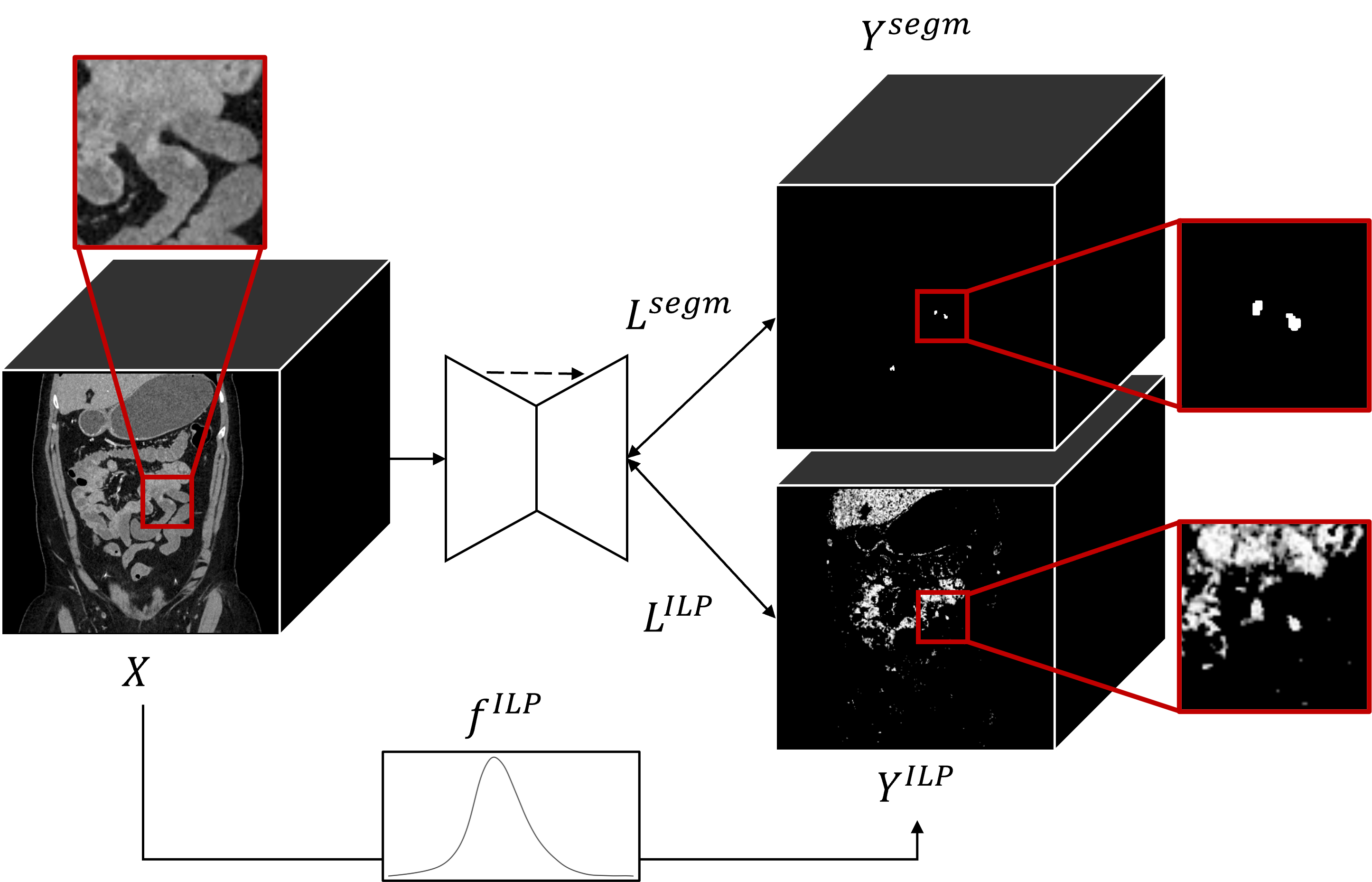}
	\caption{Augmenting a lesion segmentation network with the intensity distribution supervision. Data involved in training (examples from the small bowel carcinoid tumor dataset) are visualized. Given a network that is usually trained using pairs of an input image volume $X$ and corresponding GT segmentation $Y^{segm}$, it is augmented using an additional supervision $Y^{ILP}$, which represents probabilities of being the target lesion for each voxel according to the intensity value.
    The GT ILP map $Y^{ILP}$ for training is generated from the input volume $X$ using the ILP function $f^{ILP}$ of Figure~\ref{fig:ilp_func}, at no additional labeling cost.
    The network now predicts two outputs, namely, lesion segmentation and ILP.
    They are compared against GT labels $Y^{segm}$ and $Y^{ILP}$ to compute their respective losses $L^{segm}$ and $L^{ILP}$.
    }
	\label{fig:net_segm}
\end{figure}

The resulting ILP functions are superimposed with their corresponding histograms in Figure~\ref{fig:ilp_func}. They enable faster calculation of the ILP for a large set of voxels (a whole CT scan) than using the histogram. The ILP function, $f^{ILP}$, is used to compute the probability of being part of the target lesion for each voxel according to its intensity value. Given an input image volume $X = \{x_{i}\}_{i=1}^{N}$, the corresponding ILP volume $Y^{ILP}$ is defined as:
\begin{equation}
    \label{eq:y_il}
    Y^{ILP} = \{y_{i}^{ILP}\}_{i=1}^{N} = \{f^{ILP}(x_{i})\}_{i=1}^{N},
\end{equation}
where $N$ is the number of voxels. An example of the computed ILP volume is visualized in Figure~\ref{fig:net_segm}.
It is then provided to a network as the label map of an auxiliary task. It informs the network about our region of interest, which could contain target lesions, especially in terms of intensity values. Compared to the organ segmentation trained jointly with lesion identification tasks, our new task can be understood as a soft and possibly disconnected surrogate of organ segmentation, and it requires no additional labeling effort. 

\subsection{Network Training}\label{subsec:net_training}
\subsubsection{Lesion Segmentation Network}

The proposed intensity distribution supervision can be easily used for a lesion segmentation network. Figure~\ref{fig:net_segm} visualizes the data used for the network training.
Given an input image volume $X$, the corresponding ILP volume $Y^{ILP}$ is generated using the ILP function $f^{ILP}$.
Then, it is used as supervision for network training together with the segmentation GT, $Y^{segm}$.

For simplicity, we use a network with two output channels, which is similar to the one for joint liver and liver tumor segmentation~\citep{tang20}.
However, the second output channel of our network predicts the ILP in place of organ segmentation. The generated GT ILP volume $Y^{ILP}$ is used as supervision for this channel.

A new loss term for the added task, $L^{ILP}$, is incorporated into training accordingly as shown in Figure~\ref{fig:net_segm}.
Cross-entropy loss is used to measure the dissimilarity between the GT and the prediction of the ILP.
Finally, the overall loss function for training the lesion segmentation network is defined as:
\begin{equation}
    \label{eq:loss_net_segm}
    L = L^{segm} + \lambda L^{ILP}
\end{equation}
where $L^{segm}$ is the segmentation loss and $\lambda$ is the relative weight for the ILP loss $L^{ILP}$.
We use the generalized Dice loss~\citep{sudre17} for $L^{segm}$.

\subsubsection{Lesion Detection Network}
\begin{figure}[t]
	\centering
	\includegraphics[width = 0.7\columnwidth]{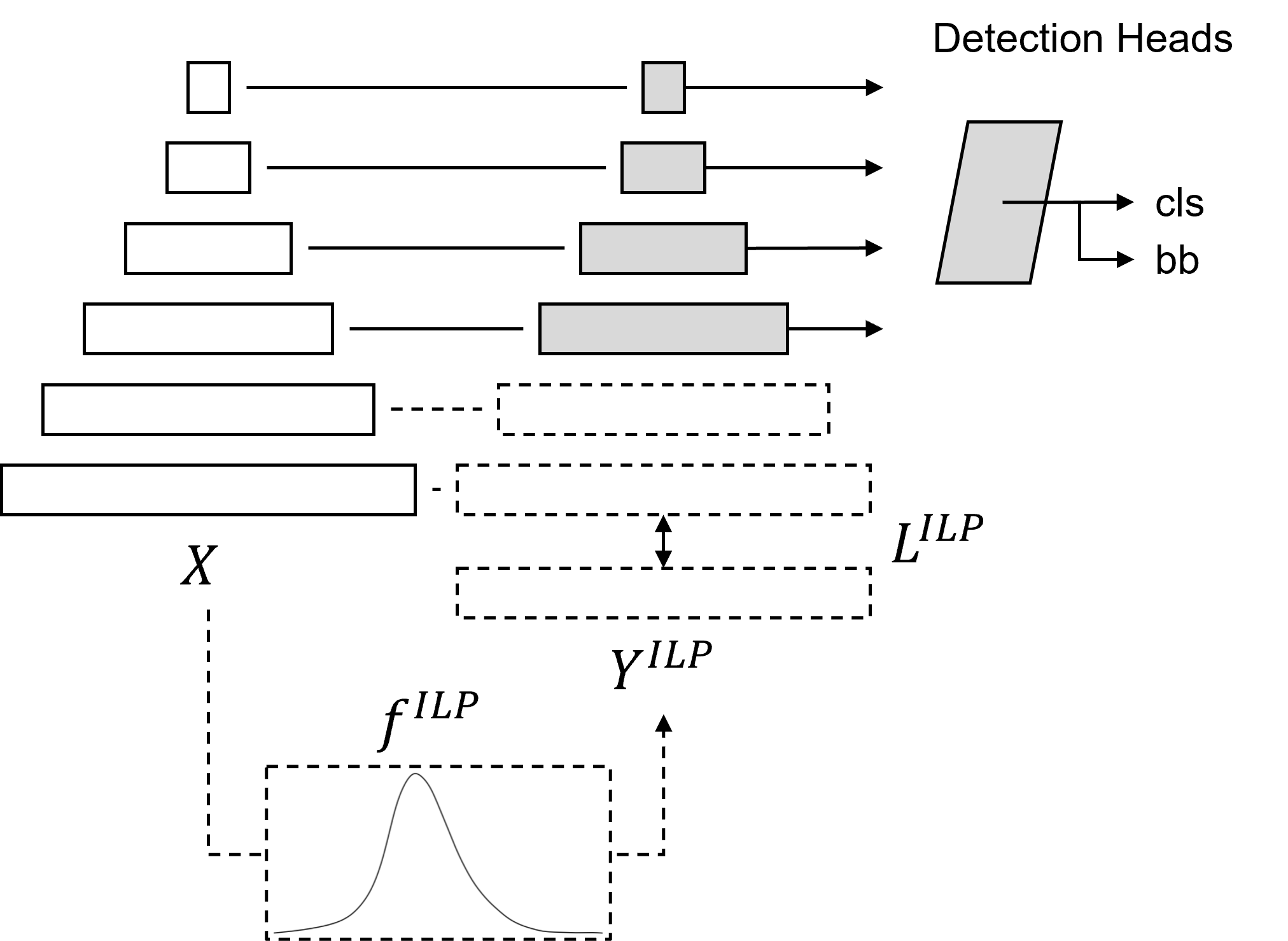}
	\caption{The use of intensity distribution supervision for lesion detection networks. The Retina Net~\citep{lin17_iccv} and Retina U-Net~\citep{jaeger20} are exemplified here. The solid line part represents the Retina Net architecture. `cls' and `bb' denote classification and box regression, respectively. The Retina U-Net is implemented by stacking additional high-resolution feature levels for the decoding part of the Retina Net and performing lesion segmentation on top of them. For this, GT segmentations are assumed to be available together with detection GTs.
    Our ILP map $Y^{ILP}$ generated from each input volume $X$ using the ILP function $f^{ILP}$ can effectively replace the GT segmentation.
    }
	\label{fig:net_det}
\end{figure}

Our intensity distribution supervision can be also used to enhance detectors that are based on feature pyramid networks (FPNs)~\citep{lin17_cvpr}, such as the Retina Net~\citep{lin17_iccv}, with the same philosophy as the Retina U-Net~\citep{jaeger20}. Figure~\ref{fig:net_det} explains the concept. In the Retina U-Net, to exploit available GT segmentation of lesions, the decoding part of the Retina Net is augmented by additional high-resolution feature levels, and semantic lesion segmentation is performed on top of them. Despite its effectiveness, it is not feasible if the GT segmentation is unavailable.

Our ILP map $Y^{ILP}$, which is generated from each input image volume $X$ using the ILP function $f^{ILP}$ can replace the GT segmentation. The same ILP loss $L^{ILP}$ is applied as in the segmentation network. Finally, the overall loss function for training the lesion detection network is defined as:
\begin{equation}
    \label{eq:loss_net_det}
    L = L^{det} + \lambda L^{ILP}
\end{equation}
where $L^{det}$ is the typical detection loss for classification and box regression, and $\lambda$ is the relative weight for $L^{ILP}$.

\subsection{Evaluation Details}
\subsubsection{Lesion Segmentation}

We first used our own version of the 3D U-Net~\citep{cicek16} to have more control over the training/test procedures and thus verify the pure impact of using the proposed intensity distribution supervision. Then, we further attempted to combine it with the self-configuring nnU-Net~\citep{isensee21} to achieve more optimized performance. Within this framework, the `3D' full resolution `U-Net' was used again but with higher complexity in terms of the network size, data augmentation, and test time method. We note that the proposed method of using intensity information can be used for any other segmentation networks.

The ILP functions in Figure~\ref{fig:ilp_func} were used for each dataset.
Especially for the LNDb dataset, we used the function of \emph{non-solid} nodules (Figure~\ref{fig:ilp_func}(c)) to emphasize them more during training since our goal is to improve the segmentation of them. Their distribution is wider than that of small bowel carcinoid tumors or kidney tumors because they are located in the lung parenchyma. Nevertheless, it is distinguishable from that of \emph{solid} nodules (Figure~\ref{fig:ilp_func}(d)).

Hyperparameters related to each method and each dataset are summarized in Table~\ref{tab:hyperparam_segm}. The learning rates and $\lambda$ were chosen through the grid search for both methods. While the other values were chosen through the grid search again by ourselves for the 3D U-Net, they were chosen automatically for the self-configuring nnU-Net.
We used the AdamW optimizer~\citep{loshchilov19} for the 3D U-Net. The SGD with a momentum of $0.99$ was used for the nnU-Net. In all of the implemented networks, 3x3x3 convolution kernels are used except 1x1x1 kernels for the final inference layer.

\begin{table}[t]
\centering
\setlength{\tabcolsep}{3pt}
\scriptsize
\begin{tabular}{@{}lccccc@{}}
\toprule
Method & \multicolumn{3}{c}{3D U-Net~\citep{cicek16}} & \multicolumn{2}{c}{nnU-Net~\citep{isensee21}} \\
\cmidrule(lr){1-1}\cmidrule(lr){2-4}\cmidrule(lr){5-6}
Dataset & SBCT & KiTS21 & LNDb & KiTS21 & LNDb \\
\midrule
Learning rate & $3 \times 10^{-4}$ & $10^{-3}$ & $3 \times 10^{-5}$ & $10^{-2}$ & $10^{-3}$ \\
Weight decay & $5 \times 10^{-4}$ & $5 \times 10^{-4}$ & $5 \times 10^{-4}$ & $3 \times 10^{-5}$ & $3 \times 10^{-5}$ \\
$\lambda$ & $1$ & $0.1$ & $0.01$ & $0.1$ & $0.003$ \\
\# channels & \{8, 16, 32, 64\} & \{32, 64, 128, 256\} & \{8, 16, 32, 64\} & \multicolumn{2}{c}{\{32, 64, 128, 256, 320, 320\}} \\
Spacing & $1 \times 1 \times 1$ mm$^3$ & $2 \times 2 \times 2$ mm$^3$ & $1 \times 1 \times 1$ mm$^3$ & $1 \times 1 \times 1$ mm$^3$ & $1 \times 1 \times 1$ mm$^3$ \\
Patch size & $224 \times 224 \times 224$ & $112 \times 112 \times 112$ & $224 \times 224 \times 224$ & $160 \times 112 \times 128$ & $128 \times 128 \times 128$ \\
Batch size & $1$ & $1$ & $1$ & $2$ & $2$ \\
\bottomrule
\end{tabular}
\caption{Summarization of the hyperparameters related to the lesion segmentation networks for each dataset.
$\lambda$ is the relative weight for $L^{ILP}$ in Eq.~\ref{eq:loss_net_segm}.
For each dataset, training is conducted using image patches of the provided size, which are sampled from resampled image volumes that have the provided resolution (spacing). Refer to the text for an explanation of how they were chosen.}\label{tab:hyperparam_segm}
\end{table}

For data augmentation, various geometric and photometric augmentation methods that are available in their implementation (https://github.com/MIC-DKFZ/nnUNet) were used as is for the nnU-Net. The whole set of photometric augmentations was turned on or off in its entirety to check their relevance in each dataset. Meanwhile, selective sets of augmentations were used for the 3D U-Net after performing an investigation on the effect of each method for each dataset. While only image rotation was used for the SBCT and the LNDb datasets, image scaling and elastic deformations were used as well for the KiTS21 dataset. In test time, a test time augmentation method of image mirroring was used for the nnU-Net.

For evaluation, we use per case and per lesion Dice scores. The per case Dice score denotes an average Dice score per scan. In calculating the per lesion Dice scores, tight local image volumes around each tumor were taken into account. Paired t-tests are conducted to show the statistical significance of the proposed method. We used an NVIDIA Tesla V100 32GB GPU to conduct experiments.

\subsubsection{Lesion Detection}
For all compared methods, the same backbone FPN~\citep{lin17_cvpr} based on a ResNet50~\citep{he16} was used. We used the Adam optimizer~\citep{kingma15} with a learning rate of $10^{-4}$. $0.003$ was used for $\lambda$ of Eq.~\ref{eq:loss_net_det}. The training was conducted using image patches of size $96 \times 96 \times 64$, which were sampled from scans that have isotropic voxels of $2 \times 2 \times 2$ mm$^3$. The batch size of $8$ was used. For data augmentation, image scaling, rotation, mirroring, and elastic deformations were used.

For experiments, we used the KiTS21 dataset, which has the biggest number of scans. We report average precision (AP) with an intersection over union threshold of 0.1, following the method of \citet{jaeger20}.

\section{Results}\label{sec:results}
\subsection{Lesion Segmentation}
\subsubsection{Experiments on the SBCT Dataset}
\paragraph{Quantitative Results}

Table~\ref{tab:quan_res_sbct} presents quantitative results of segmentation methods, which differ in the ways of using the intensity distribution information. We used the 3D U-Net~\citep{cicek16} to verify the pure effect of the different usages of the intensity distribution information.

\begin{table}[t]
\centering
\setlength{\tabcolsep}{3pt}
\scriptsize
\begin{tabular}{@{}lcccccc@{}}
\toprule
\multicolumn{1}{c}{\multirow{2}{*}{Method}} & \multicolumn{2}{c}{Per case} & \multicolumn{2}{c}{Per lesion} & \multicolumn{2}{c}{Per lesion ($\geq$ 125 mm$^3$)} \\
\cmidrule(lr){2-3}\cmidrule(lr){4-5}\cmidrule(lr){6-7}
& Dice (\%) & p-value & Dice (\%) & p-value & Dice (\%) & p-value \\
\midrule
3D U-Net~\citep{cicek16} & 41.3 $\pm$ 27.2 & 0.0022 & 30.0 $\pm$ 36.7 & 0.0398 & 37.7 $\pm$ 36.4 & 0.1002 \\
3D U-Net + PP & 36.2 $\pm$ 25.3 & 0.0001 & 24.7 $\pm$ 32.5 & 0.0005 & 30.5 $\pm$ 31.3 & 0.0026 \\
3D U-Net + ILP(in) & 41.6 $\pm$ 29.2 & 0.0860 & 32.8 $\pm$ 39.0 & 0.1885 & 37.9 $\pm$ 38.3 & 0.2133 \\
3D U-Net + ILP & \textbf{47.8} $\pm$ 29.6 & - & \textbf{35.9} $\pm$ 40.0 & - & \textbf{42.6} $\pm$ 39.5 & - \\
3D U-Net + ILP(shifted) & 41.1 $\pm$ 30.4 & 0.0084 & 30.1 $\pm$ 38.3 & 0.0296 & 32.4 $\pm$ 37.7 & 0.0025 \\
\bottomrule
\end{tabular}
\caption{Results of segmentation methods that use the ILP in different ways on the SBCT dataset. `3D U-Net' denotes the baseline method that performs only lesion segmentation; `3D U-Net + PP' denotes applying post-processing (PP) that is based on the ILP to the results of `3D U-Net'; `3D U-Net + ILP(in)' denotes using the ILP volume as an additional input channel instead of as an additional supervision; `3D U-Net + ILP' denotes the proposed method; `3D U-Net + ILP(shifted)' denotes the proposed method but using a shifted ILP function, which would be irrelevant with the target tumor. Dice scores were calculated at two different subject levels, namely, per case and per lesion. Refer to the text for an explanation of each of the metrics. Mean and standard deviation values are presented together. P-values are computed by conducting paired t-tests between the proposed method and the others with the Dice scores.}\label{tab:quan_res_sbct}
\end{table}

Applying post-processing to the prediction of the segmentation network, where the ILP volume $Y^{ILP}$ is multiplied with the network predicted probability map, rather worsened the performance (‘3D U-Net + PP’). 
This post-processing could oversimply rule out lesions that have intensity values deviating from the built intensity distribution.
We also tried using the ILP volume as an additional input channel instead of as additional supervision (‘3D U-Net + ILP(in)’).
It can be another way to highlight our region of interest at the input level. However, it performed merely on par with the baseline that does not use this additional information.

On the other hand, the proposed method, ‘3D U-Net + ILP’, showed clear improvements for all types of Dice scores when compared to the baseline.
The proposed method of using the intensity distribution supervision does not entail any additional labeling effort.
The ILP function can be constructed and included in training by looking up already available CT scans and corresponding GT tumor segmentation.

We also investigate the effect of having the precise intensity model of a target.
‘3D U-Net + ILP(shifted)’ is the proposed method but uses another ILP function that is +100 shifted from the original one.
The shifted function does not reflect the actual intensity distribution of the target anymore. It performed rather worse than the baseline.

All methods including the proposed method showed higher Dice scores for relatively larger tumors ($\geq$ 125 mm$^3$, which is approximately $\geq$ 6 mm diameter) than for all tumors.

\begin{figure}[t]
	\centering
    \begin{minipage}{1\columnwidth}
        \subfloat{\includegraphics[width = 0.19\columnwidth]{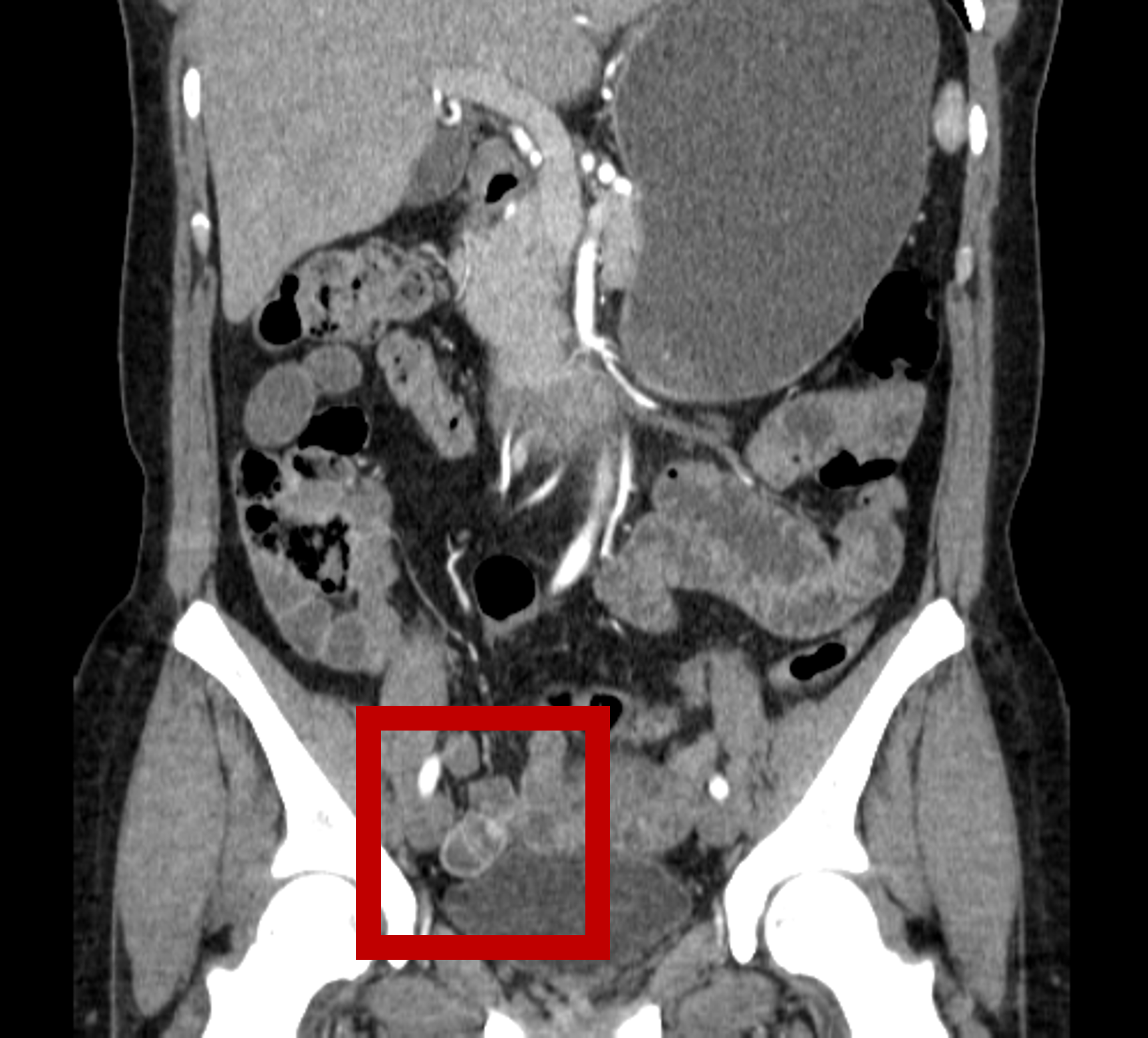}}
        \hspace{0.0001\columnwidth}
	    \subfloat{\includegraphics[width = 0.19\columnwidth]{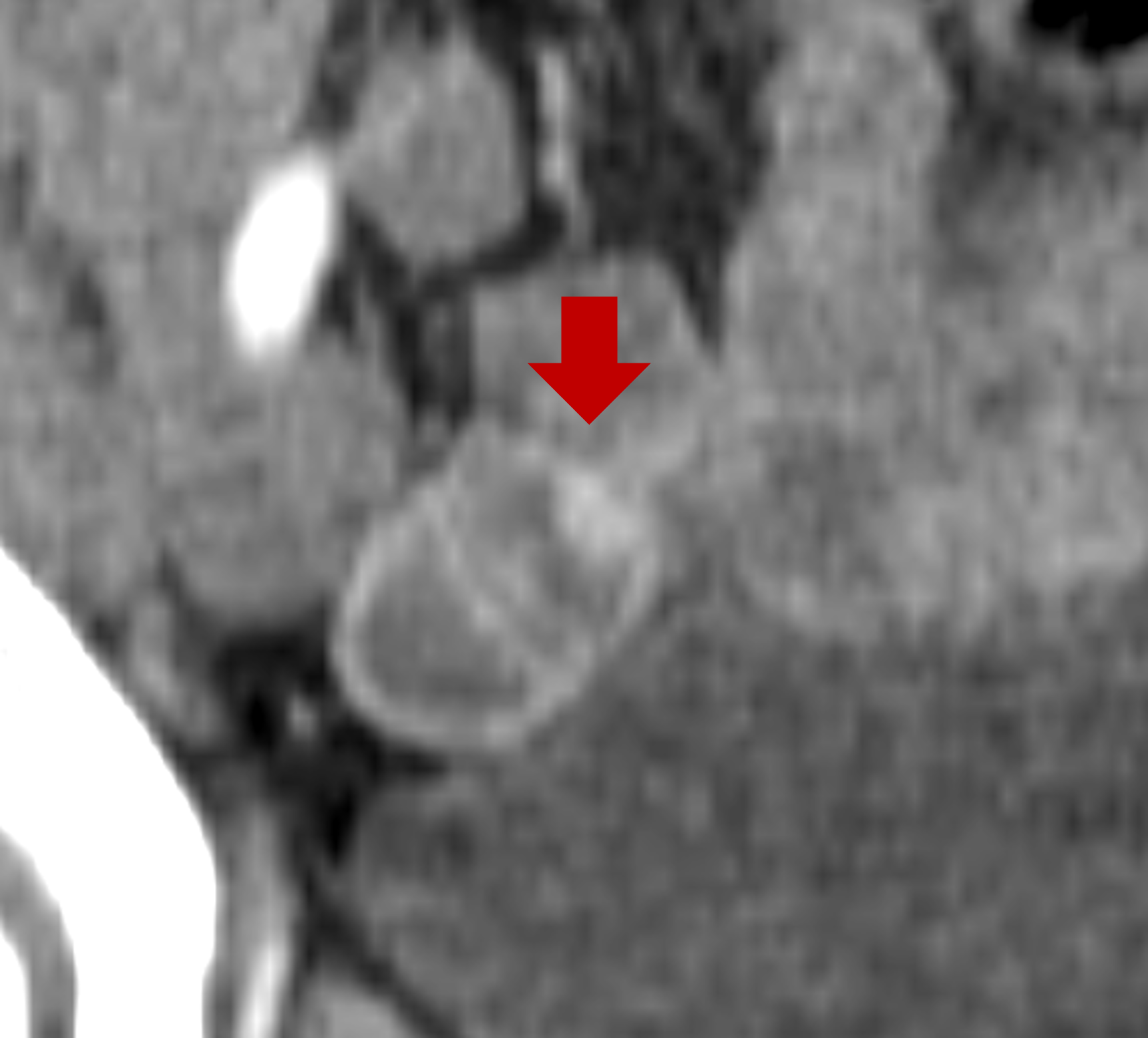}}
        \hspace{0.0001\columnwidth}
	    \subfloat{\includegraphics[width = 0.19\columnwidth]{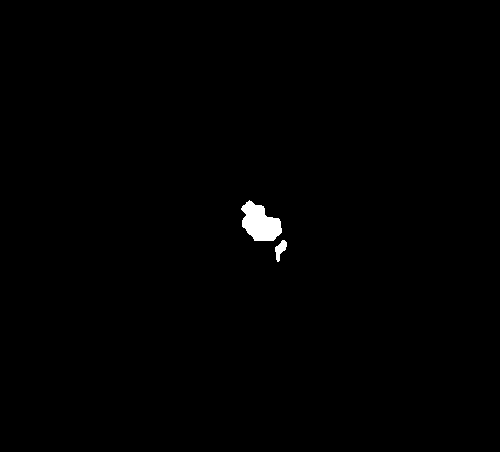}}
        \hspace{0.0001\columnwidth}
	    \subfloat{\includegraphics[width = 0.19\columnwidth]{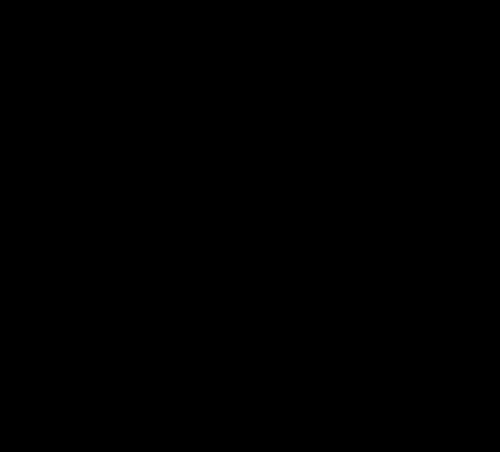}}
        \hspace{0.0001\columnwidth}
	    \subfloat{\includegraphics[width = 0.19\columnwidth]{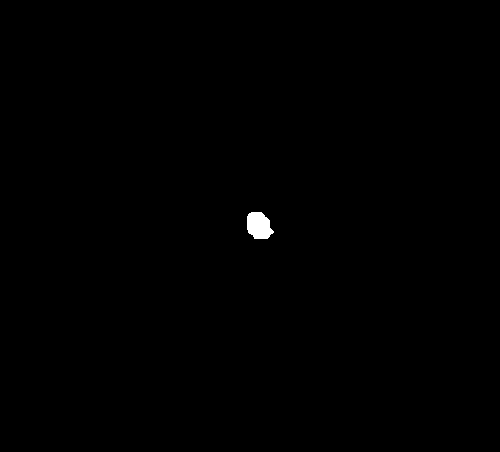}}
    \end{minipage}
    \begin{minipage}{1\columnwidth}
        \subfloat{\includegraphics[width = 0.19\columnwidth]{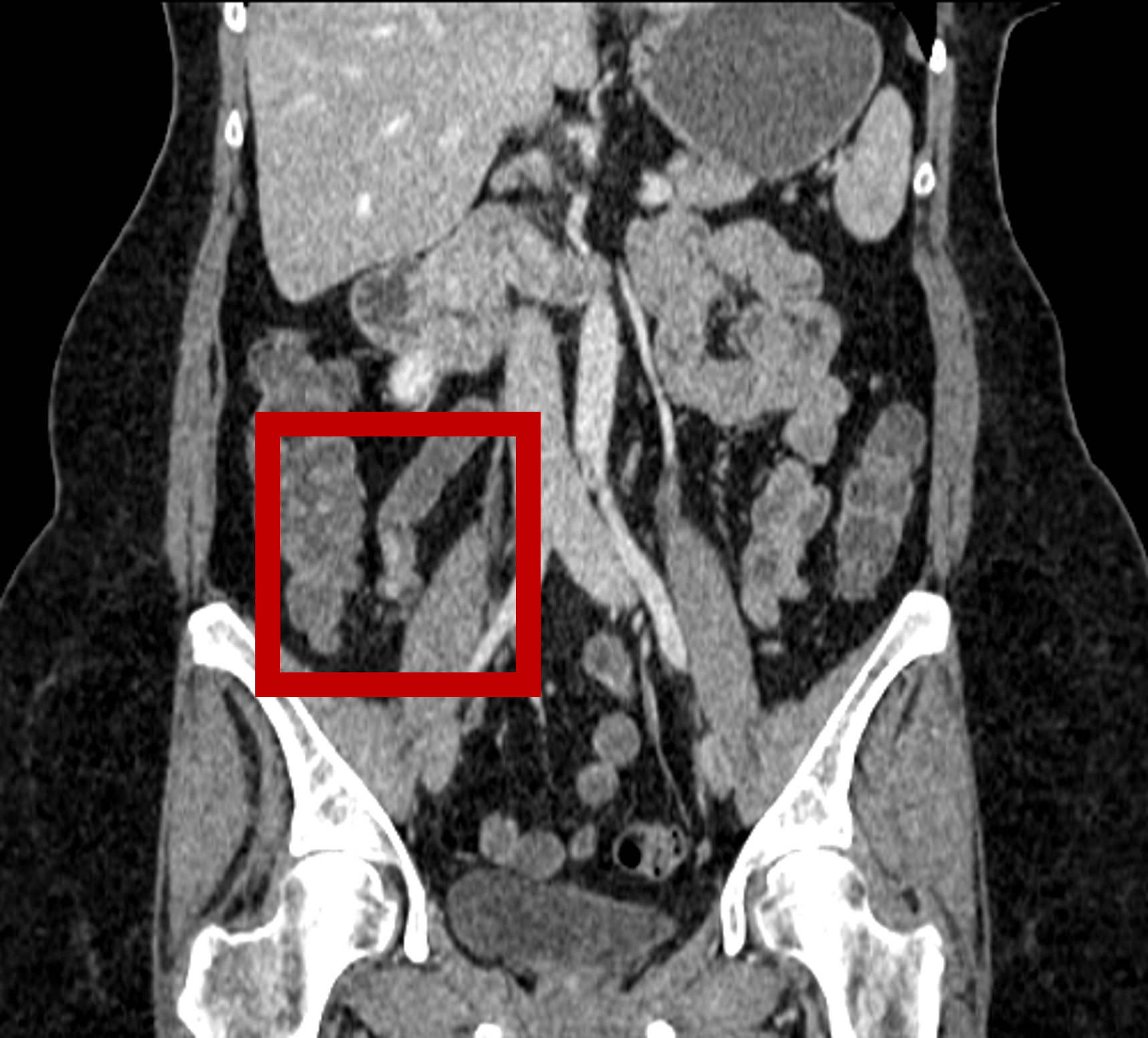}}
        \hspace{0.0001\columnwidth}
	    \subfloat{\includegraphics[width = 0.19\columnwidth]{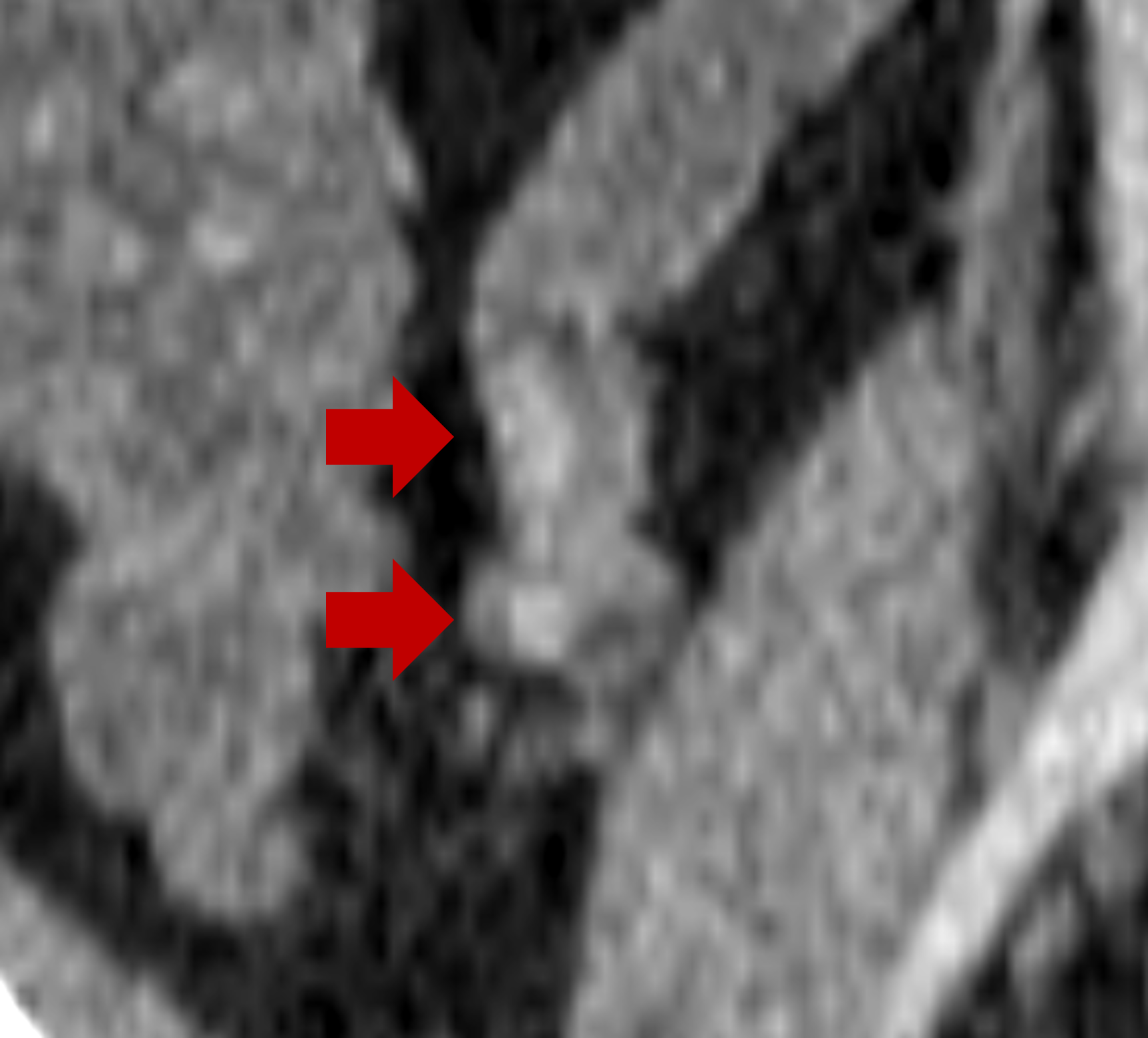}}
        \hspace{0.0001\columnwidth}
	    \subfloat{\includegraphics[width = 0.19\columnwidth]{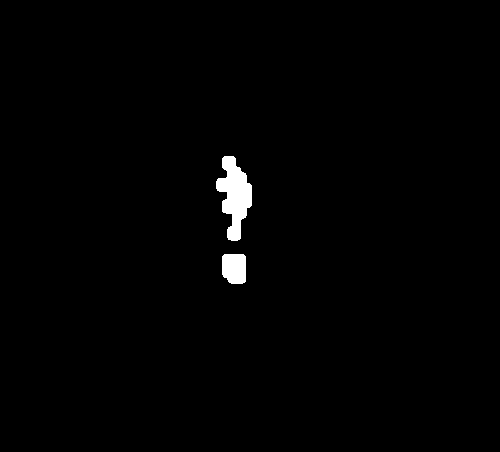}}
        \hspace{0.0001\columnwidth}
	    \subfloat{\includegraphics[width = 0.19\columnwidth]{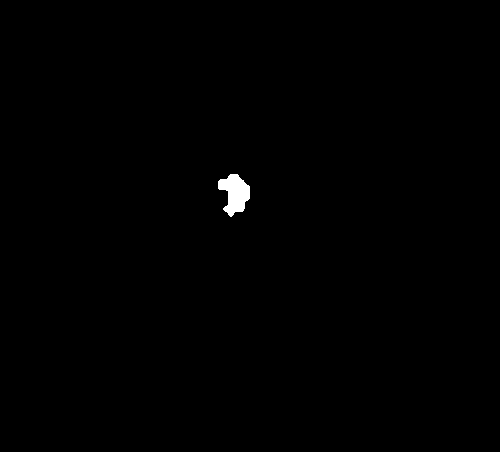}}
        \hspace{0.0001\columnwidth}
	    \subfloat{\includegraphics[width = 0.19\columnwidth]{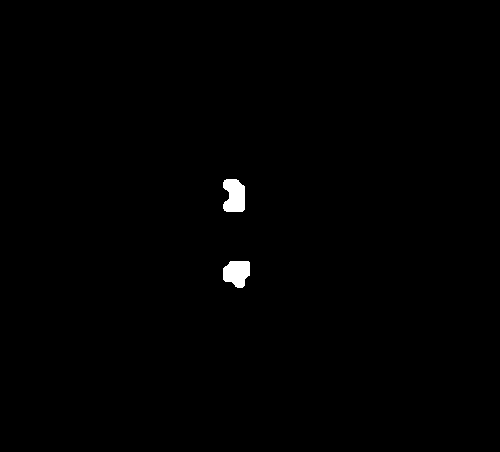}}
    \end{minipage}
    \begin{minipage}{1\columnwidth}
        \subfloat{\includegraphics[width = 0.19\columnwidth]{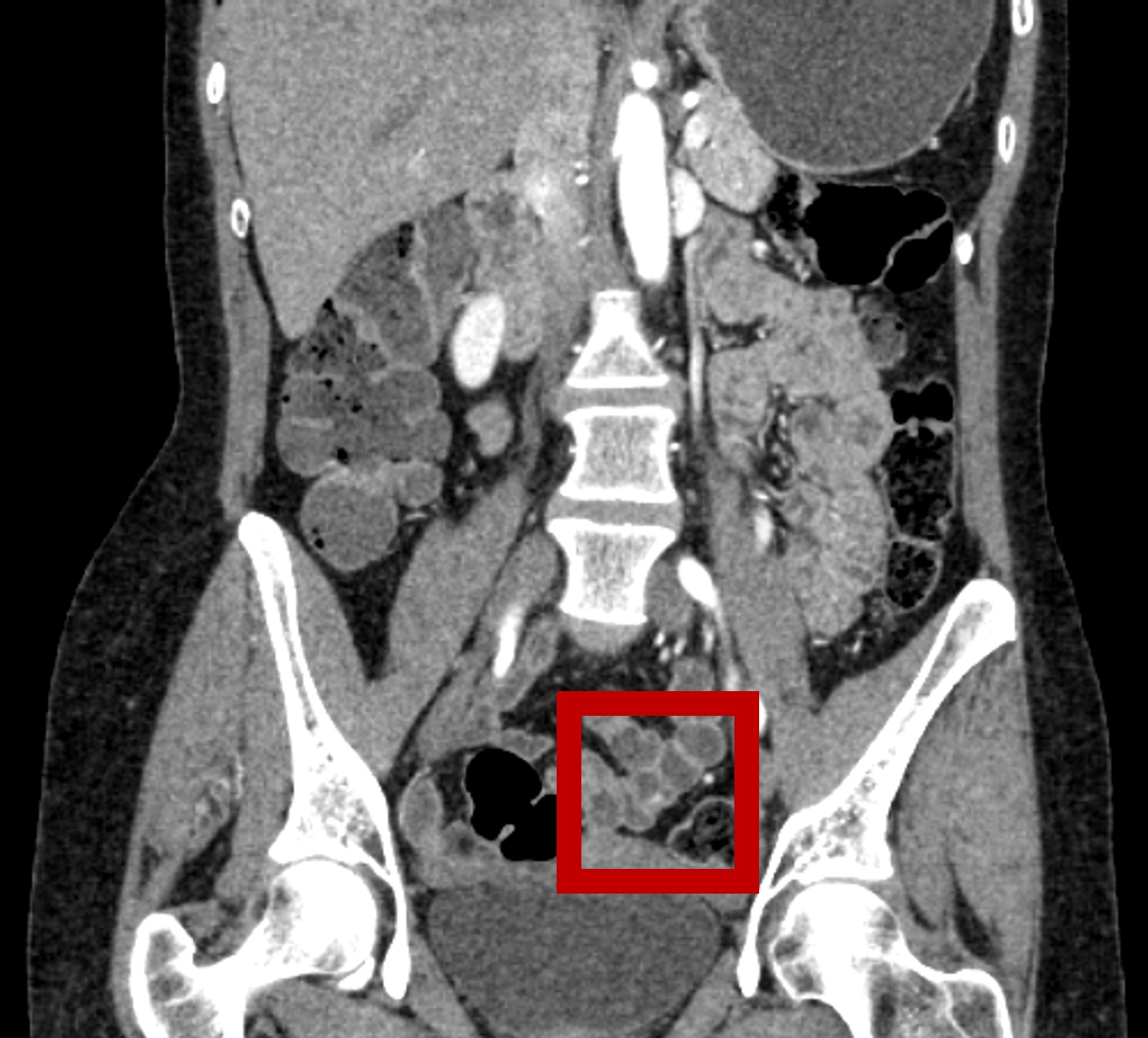}}
        \hspace{0.0001\columnwidth}
	    \subfloat{\includegraphics[width = 0.19\columnwidth]{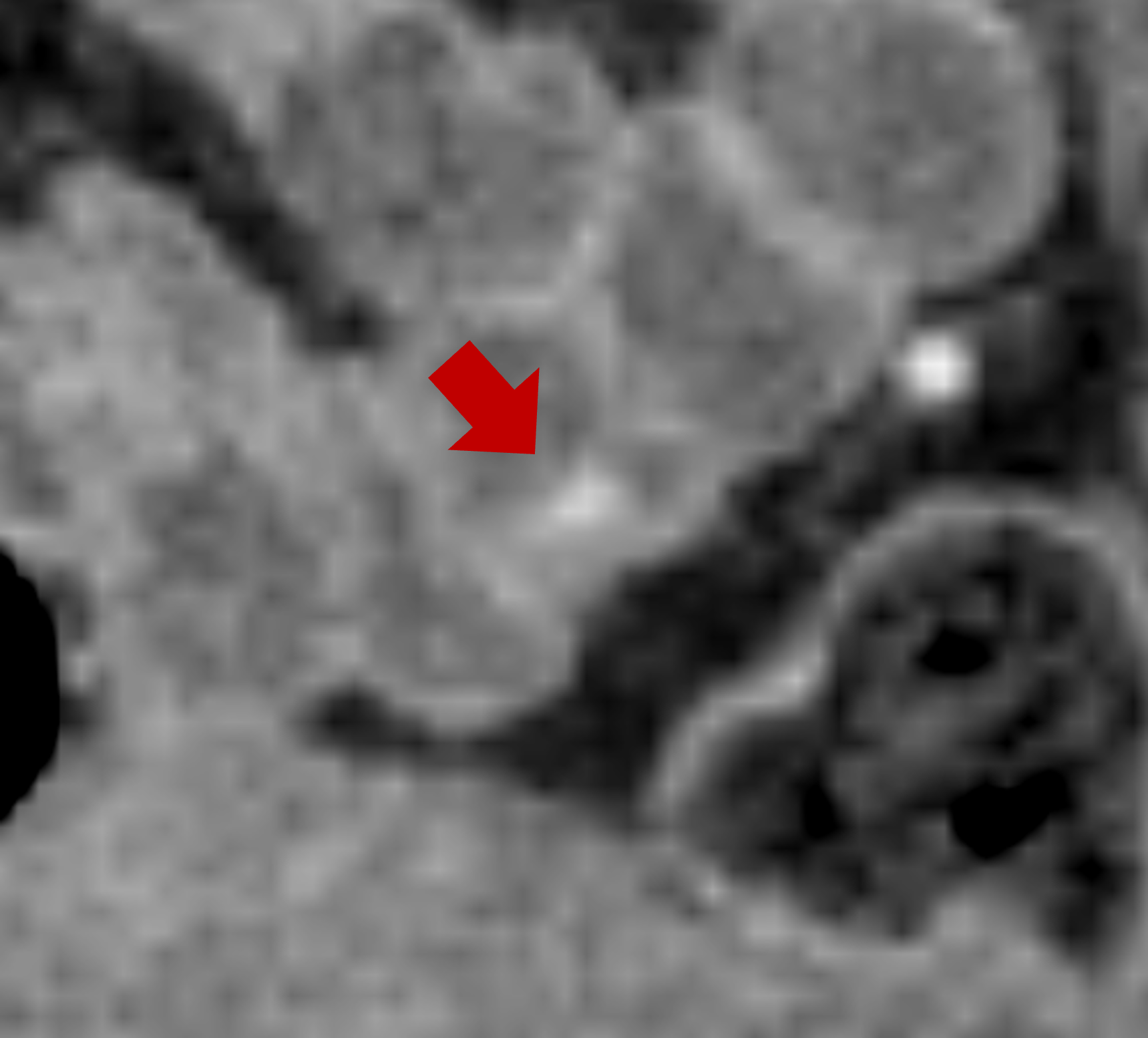}}
        \hspace{0.0001\columnwidth}
	    \subfloat{\includegraphics[width = 0.19\columnwidth]{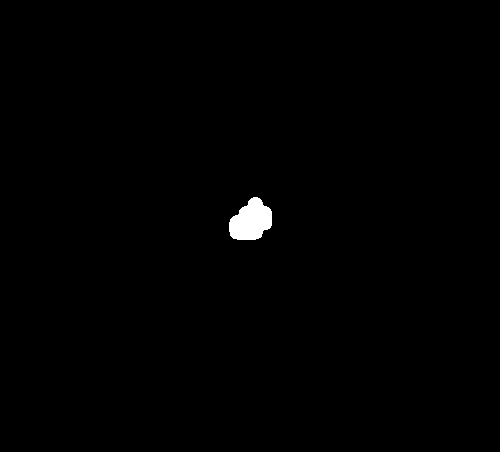}}
        \hspace{0.0001\columnwidth}
	    \subfloat{\includegraphics[width = 0.19\columnwidth]{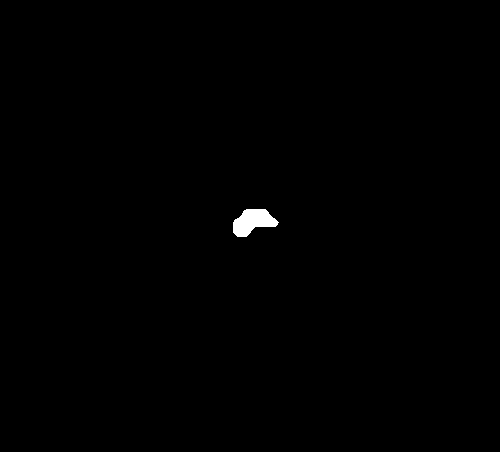}}
        \hspace{0.0001\columnwidth}
	    \subfloat{\includegraphics[width = 0.19\columnwidth]{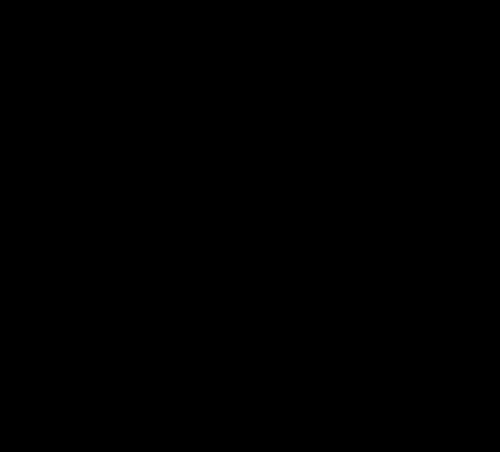}}
    \end{minipage}
     \caption{Example segmentation results on the SBCT dataset. Each row represents different cases. The columns, from left, represent input CT scan, zoomed view of the red box in the CT scan (carcinoid tumors are pointed by the red arrows), corresponding GT segmentation, result of the baseline method (`3D U-Net' in Table~\ref{tab:quan_res_sbct}), and result of the proposed method (`3D U-Net + ILP' in Table~\ref{tab:quan_res_sbct}), respectively.}
	\label{fig:qual_res_sbct}
\end{figure}

\paragraph{Qualitative Results}

Figure~\ref{fig:qual_res_sbct} presents example segmentation results of small bowel carcinoid tumor. Compared to the baseline method that is trained without the intensity distribution supervision, the proposed method segments more tumors (first and second rows). The last row shows a failure case, where the proposed method missed a blurry small tumor.

\subsubsection{Experiments on the KiTS21 Dataset}
\paragraph{Quantitative Results}

Table~\ref{tab:quan_res_kits21_segm} presents quantitative results of different segmentation methods on the KiTS21 dataset. We first used different versions of the 3D U-Net that were augmented using different additional supervision, and again used the nnU-Net for more optimized performance.
We also incorporated the proposed intensity distribution supervision in training the winning method of the KiTS21 challenge~\citep{zhao21}.

\begin{table}[t]
\centering
\setlength{\tabcolsep}{3pt}
\begin{tabular}{@{}lcccc@{}}
\toprule
\multicolumn{1}{c}{\multirow{2}{*}{Method}} & \multicolumn{2}{c}{Per case} & \multicolumn{2}{c}{Per lesion} \\
\cmidrule(lr){2-3}\cmidrule(lr){4-5}
& Dice (\%) & p-value & Dice (\%) & p-value \\
\midrule
3D U-Net~\citep{cicek16} & 62.0 $\pm$ 34.8 & 0.0216 & 66.7 $\pm$ 36.0 & 0.0934 \\
3D U-Net + organ & 68.3 $\pm$ 30.5 & 0.3713 & \textbf{71.6} $\pm$ 31.4 & 0.5727 \\
3D U-Net + ILP & \textbf{69.2} $\pm$ 28.5 & - & 71.2 $\pm$ 30.8 & - \\
\midrule
nnU-Net~\citep{isensee21} & 74.2 $\pm$ 26.8 & 0.1494 & 77.6 $\pm$ 27.1 & 0.1742 \\
nnU-Net + photo aug. & 74.2 $\pm$ 29.8 & 0.1072 & 76.9 $\pm$ 31.0 & 0.0887 \\
nnU-Net + organ & \textbf{80.4} $\pm$ 23.7 & 0.9857 & \textbf{80.3} $\pm$ 27.4 & 0.7983 \\
nnU-Net + ILP & 76.0 $\pm$ 27.1 & - & 79.1 $\pm$ 27.9 & - \\
\midrule
\citet{zhao21} &  74.9 $\pm$ 27.0 & 0.1264 & 81.3 $\pm$ 22.3 & 0.1716 \\
\citet{zhao21} + ILP & \textbf{77.0} $\pm$ 23.3 & - & \textbf{83.0} $\pm$ 18.1 & - \\
\bottomrule
\end{tabular}
\caption{Results of different segmentation methods on the KiTS21 dataset. The first three and the next four are based on the 3D U-Net and the nnU-Net, respectively.
The last two are based on the winning method of the KiTS21 challenge.
For the self-configuring nnU-Net method, the `3D' full resolution `U-Net' is used again but it has a higher complexity in terms of the network size, data augmentation, and test time method, etc. `+ organ' denotes performing organ segmentation jointly with tumor segmentation;
`+ ILP' denotes the proposed method;
`+ photo aug.' denotes utilizing photometric augmentations as well as geometric augmentations in training.}\label{tab:quan_res_kits21_segm}
\end{table}

In the 3D U-Net based comparison, the proposed method (`3D U-Net + ILP') outperformed the baseline (`3D U-Net').
We further compare it against a multi-task learning network that performs organ (kidney) segmentation together with lesion (kidney tumor) segmentation, which is `3D U-Net + organ' in Table~\ref{tab:quan_res_kits21_segm}.
As our ILP map informs the network about our region of interest, which could contain target lesions, in terms of intensity values, organ segmentation supervision could do the same in a stricter way, i.e., kidney tumors can exist within the kidney.
The proposed method performed on par with the organ-segmentation-augmented method, which requires additional labeling effort while the proposed method does not.

The proposed method (`nnU-Net + ILP') still outperformed the baseline (`nnU-Net') when the nnU-Net was used.
The test time augmentation method of the nnU-Net could decrease the performance gap by benefiting an under-performed method more. We note that the proposed method could be not well harmonized with photometric augmentations since they randomly distort original voxel values and thus can change the physical meaning that each voxel originally has on CT scans. We found in this dataset that photometric augmentations do not really help in improving the performance even for the baseline method (`nnU-Net + photo aug.'). Thus, only geometric augmentations were used. `nnU-Net + organ' showed a better performance than the proposed method, but it used an additional annotation of the kidney. 

We also incorporated the proposed intensity distribution supervision in training the winning method of the KiTS21 challenge~\citep{zhao21}. The method is composed of three steps (networks), which are coarse kidney segmentation, fine kidney segmentation, and tumor segmentation.
Therefore, it uses GT segmentations of the kidney and tumor for training of the first and second networks, and the last network, respectively.
Since there is no publicly available code, we have used our own implementation for the experiment.
When the intensity distribution supervision was incorporated in the last tumor segmentation step at no additional labeling cost, a better performance was again achieved (`\citet{zhao21} + ILP').

\begin{figure}[t]
   \begin{center}
   \includegraphics[width=0.7\columnwidth]{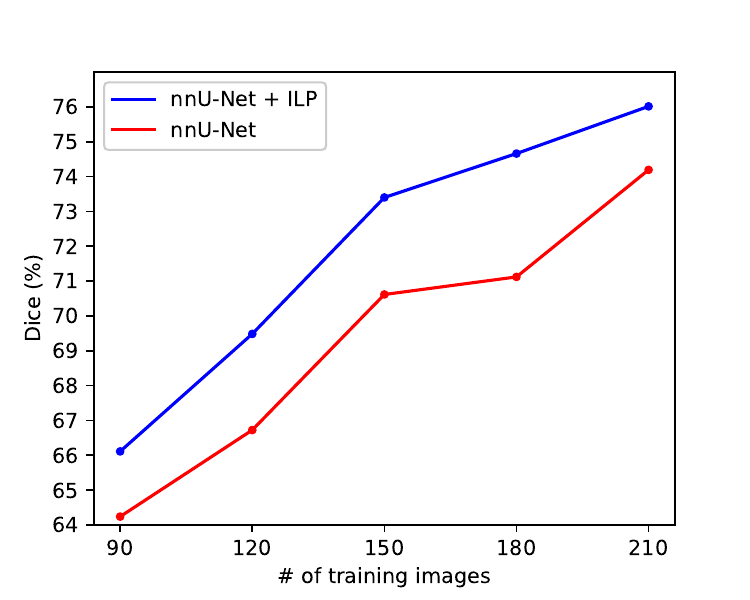}
   \caption{Segmentation performances of the baseline nnU-Net and the proposed method depending on the number of training images on the KiTS21 dataset. Per case Dice scores are reported.}
   \label{fig:quan_res_kits21_segm_by_train_size}
   \end{center}
\end{figure}

Figure~\ref{fig:quan_res_kits21_segm_by_train_size} shows the segmentation performances on the KiTS21 dataset, depending on the number of training images. Given the original training set of 210 images, 90, 120, 150, or 180 images were randomly sampled to conduct the experiments. The same validation and test sets of 30 and 60 images, respectively, were used for all training set sizes. The proposed method consistently outperformed the baseline for all experiments, with a margin of around 2\%.

\begin{figure}[t]
	\centering
    \begin{minipage}{1\columnwidth}
        \subfloat{\includegraphics[width = 0.24\columnwidth]{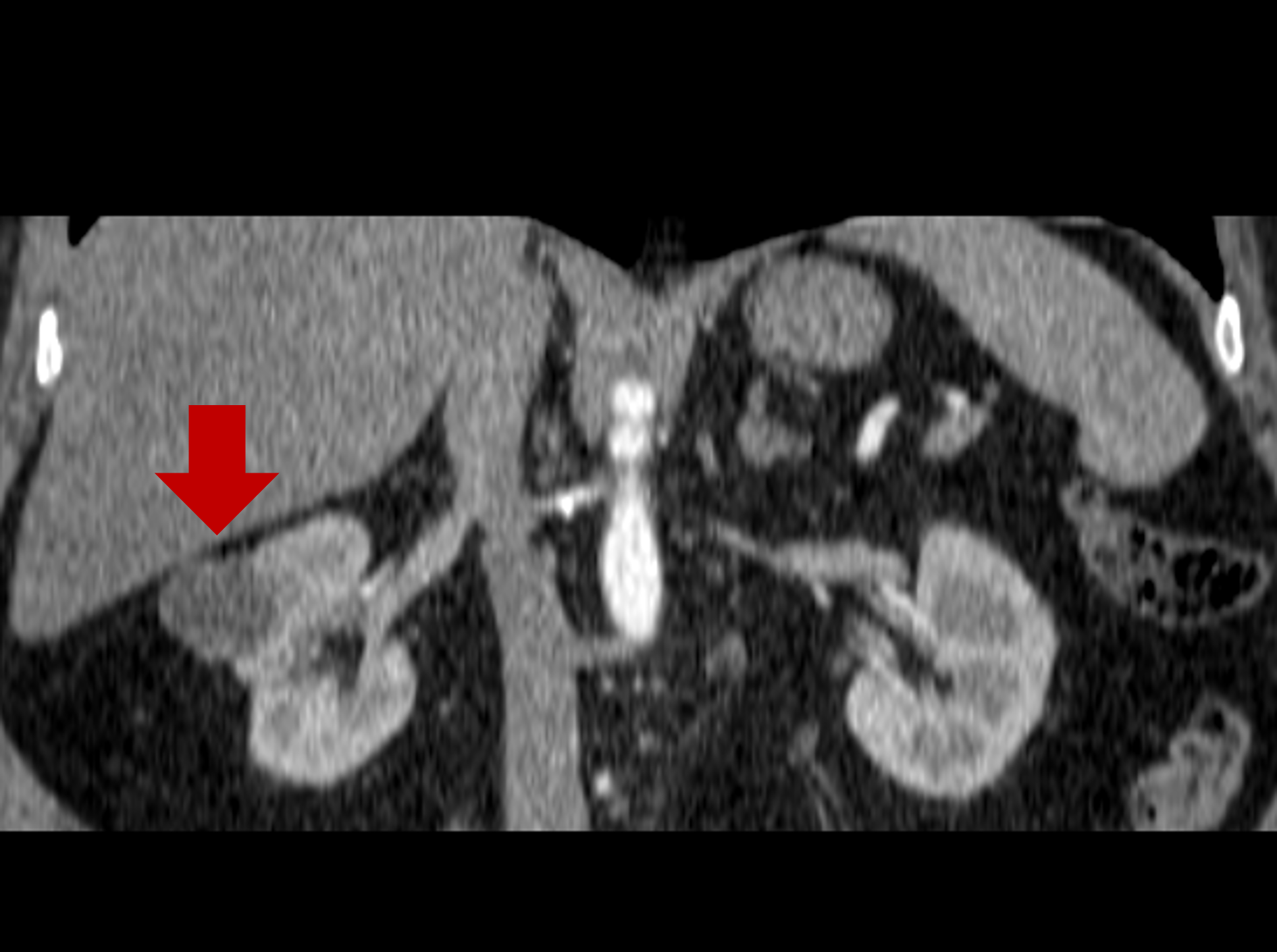}}
        \hspace{0.0001\columnwidth}
	    \subfloat{\includegraphics[width = 0.24\columnwidth]{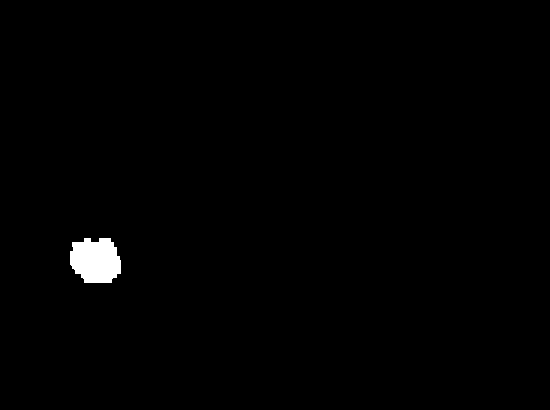}}
        \hspace{0.0001\columnwidth}
	    \subfloat{\includegraphics[width = 0.24\columnwidth]{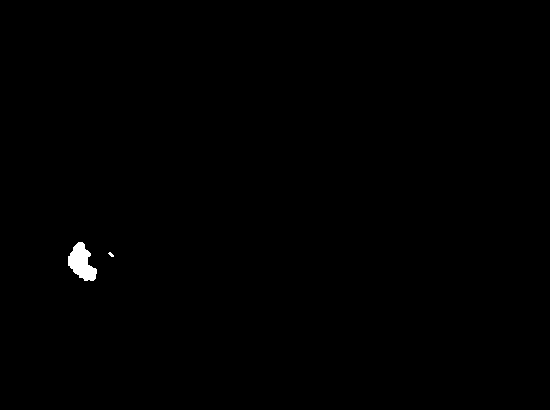}}
        \hspace{0.0001\columnwidth}
	    \subfloat{\includegraphics[width = 0.24\columnwidth]{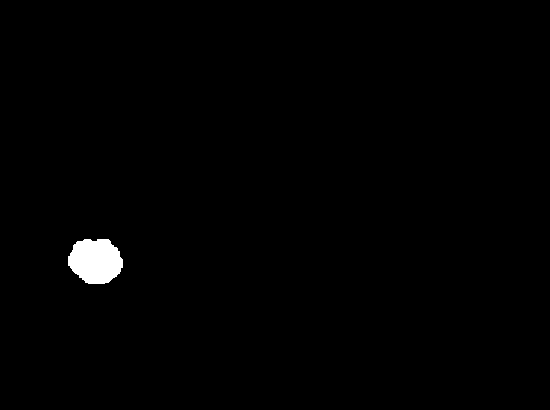}}
    \end{minipage}
    \begin{minipage}{1\columnwidth}
        \subfloat{\includegraphics[width = 0.24\columnwidth]{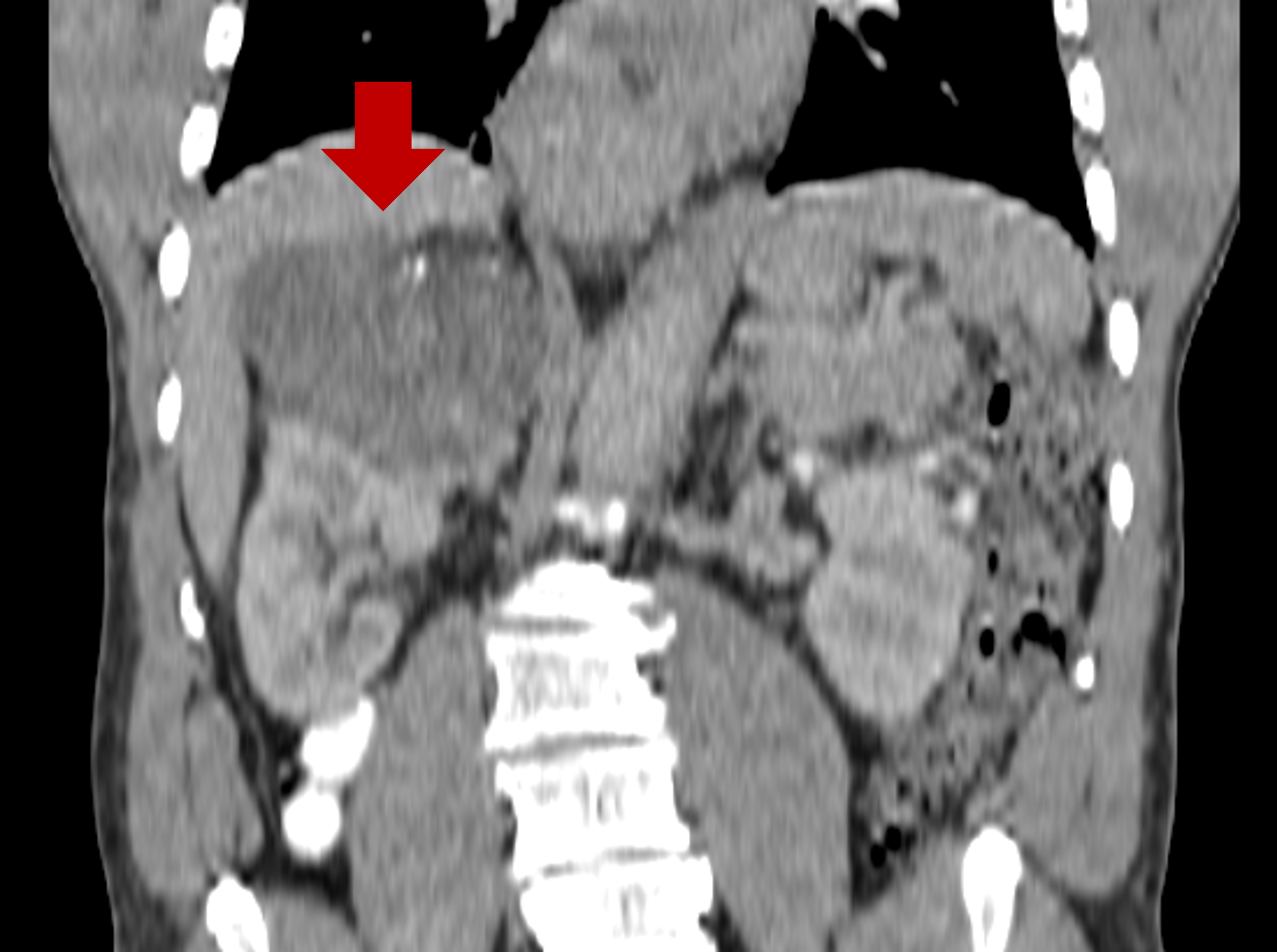}}
        \hspace{0.0001\columnwidth}
	    \subfloat{\includegraphics[width = 0.24\columnwidth]{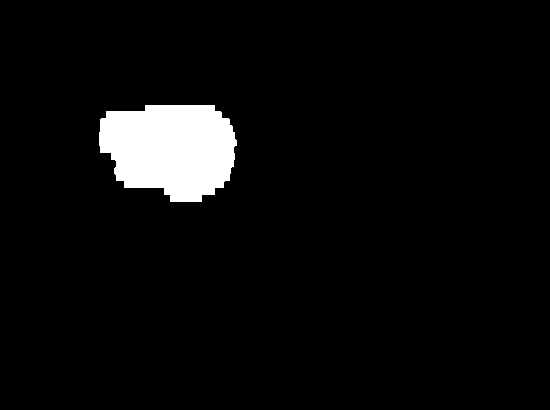}}
        \hspace{0.0001\columnwidth}
	    \subfloat{\includegraphics[width = 0.24\columnwidth]{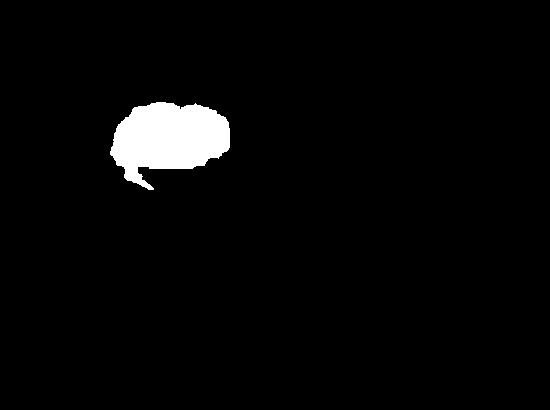}}
        \hspace{0.0001\columnwidth}
	    \subfloat{\includegraphics[width = 0.24\columnwidth]{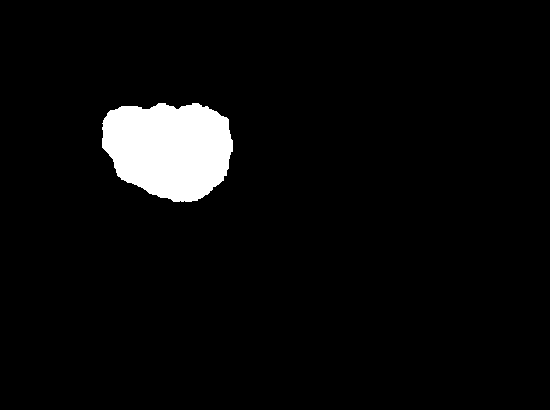}}
    \end{minipage}
    \begin{minipage}{1\columnwidth}
        \subfloat{\includegraphics[width = 0.24\columnwidth]{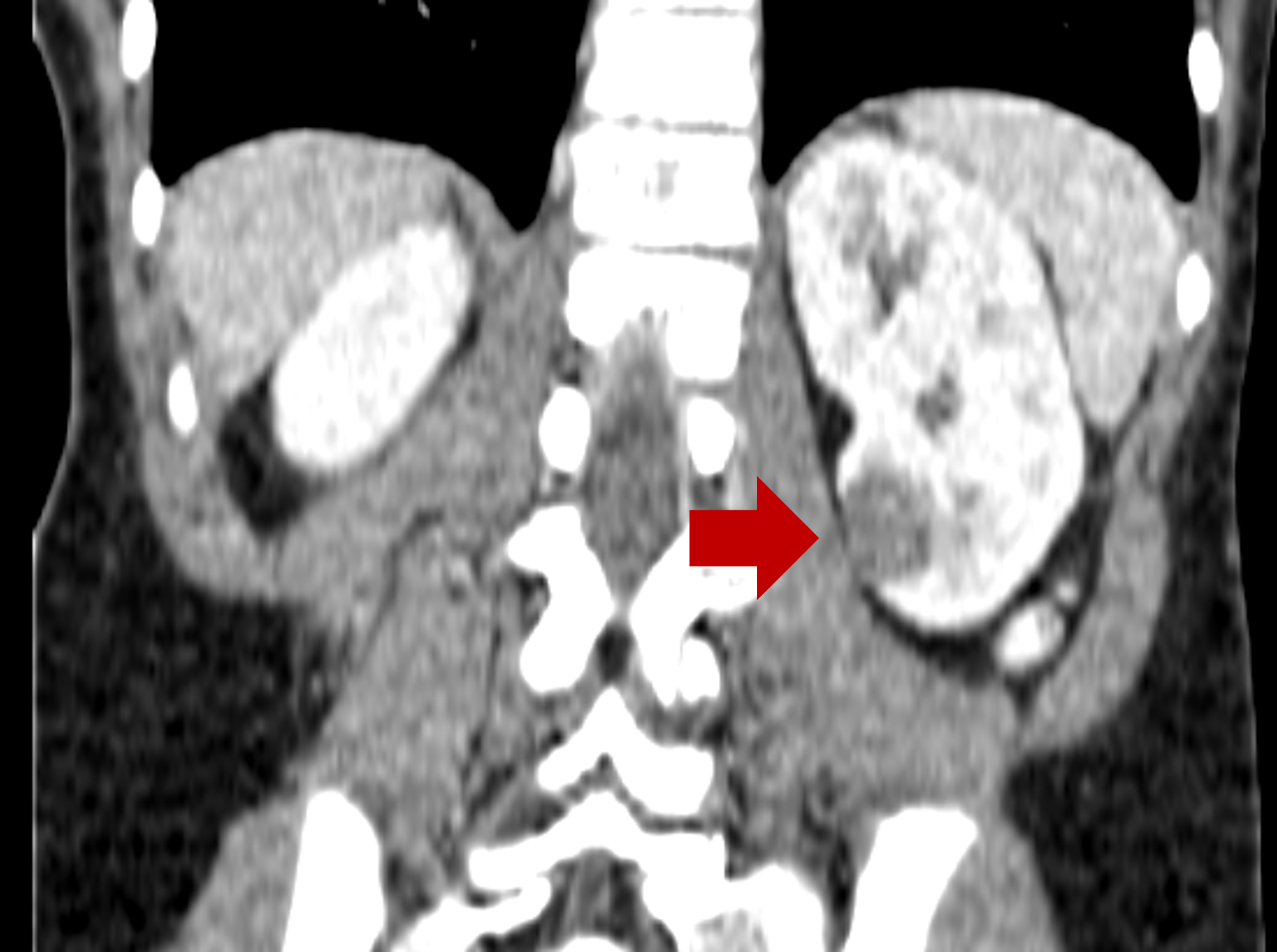}}
        \hspace{0.0001\columnwidth}
	    \subfloat{\includegraphics[width = 0.24\columnwidth]{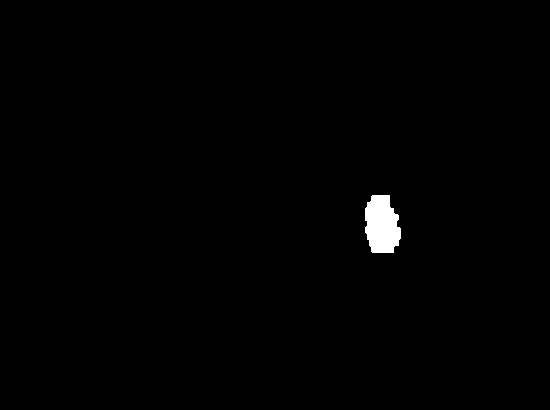}}
        \hspace{0.0001\columnwidth}
	    \subfloat{\includegraphics[width = 0.24\columnwidth]{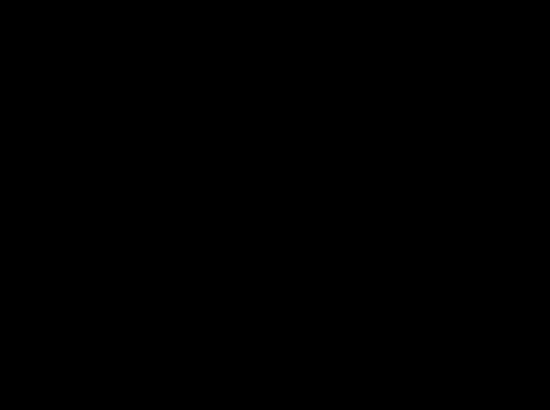}}
        \hspace{0.0001\columnwidth}
	    \subfloat{\includegraphics[width = 0.24\columnwidth]{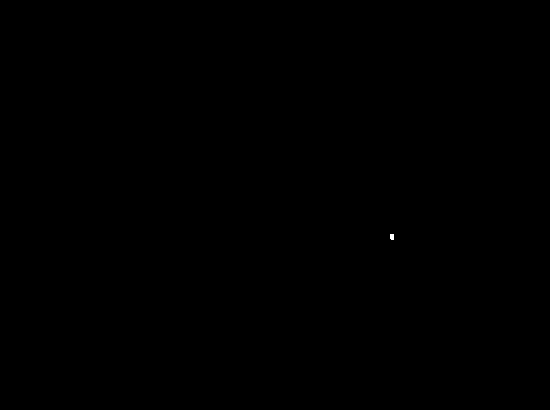}}
    \end{minipage}
    \caption{Example segmentation results on the KiTS21 dataset. Each row represents different cases. The columns, from left, represent input CT scan (kidney tumors are pointed by the red arrows), corresponding GT segmentation, result of the baseline method (`nnU-Net' in Table~\ref{tab:quan_res_kits21_segm}), and result of the proposed method (`nnU-Net + ILP' in Table~\ref{tab:quan_res_kits21_segm}), respectively.}
	\label{fig:qual_res_kits21_segm}
\end{figure}

\paragraph{Qualitative Results}
Figure~\ref{fig:qual_res_kits21_segm} shows example segmentation results on the KiTS21 dataset. The proposed method segments tumors more precisely (first and second rows) by utilizing the intensity distribution supervision when compared to the baseline.
The last row shows a failure case, where the tumor was missed by both the baseline and proposed methods.

\subsubsection{Experiments on the LNDb Dataset}
\paragraph{Quantitative Results}

\begin{table}[t]
\centering
\setlength{\tabcolsep}{3pt}
\footnotesize
\begin{tabular}{@{}lccccc@{}}
\toprule
\multicolumn{1}{c}{\multirow{2}{*}{Method}} & \multicolumn{1}{c}{\multirowcell{2}{Nodule\\ texture}} & \multicolumn{2}{c}{Per case} & \multicolumn{2}{c}{Per lesion} \\
\cmidrule(lr){3-4}\cmidrule(lr){5-6}
& & Dice (\%) & p-value & Dice (\%) & p-value \\
\midrule
\multicolumn{1}{c}{\multirow{2}{*}{3D U-Net~\citep{cicek16}}} & all & 21.6 $\pm$ 21.7 & 0.3022 & 38.1 $\pm$ 36.4 & 0.9573 \\
& non-solid & 10.9 $\pm$ 17.1 & 0.0782 & 26.0 $\pm$ 33.6 & 0.7471 \\
\multicolumn{1}{c}{\multirow{2}{*}{3D U-Net + ILP}} & all & 24.0 $\pm$ 26.7 & - & 32.5 $\pm$ 34.4 & - \\
& non-solid & 17.6 $\pm$ 24.8 & - & 22.5 $\pm$ 30.4 & - \\
\midrule
\multicolumn{1}{c}{\multirow{2}{*}{nnU-Net~\citep{isensee21}}} & all & 26.4 $\pm$ 27.4 & 0.0588 & 28.0 $\pm$ 36.5 & 0.0283 \\
& non-solid & 10.4 $\pm$ 21.9 & 0.1537 & 13.8 $\pm$ 27.4 & 0.0587 \\
\multicolumn{1}{c}{\multirow{2}{*}{nnU-Net + ILP}} & all & 32.7 $\pm$ 25.2 & - & 33.8 $\pm$ 36.6 & - \\
& non-solid & 14.0 $\pm$ 23.6 & - & 21.0 $\pm$ 31.2 & - \\
\bottomrule
\end{tabular}
\caption{Results of different segmentation methods on the LNDb dataset.
The proposed methods that use the ILP supervision (+ ILP) are compared with the baseline methods of the 3D U-Net and nnU-Net.
For each method, the performances on all nodules and on only non-solid nodules are presented.}\label{tab:quan_res_lndb}
\end{table}

Table~\ref{tab:quan_res_lndb} presents quantitative segmentation results on the LNDb dataset. As mentioned in Section~\ref{sec:datasets}, we tried to improve the segmentation of \emph{non-solid} nodules in this work by incorporating their intensity distribution information into network training.
For both the 3D U-Net and nnU-Net, the inclusion of the intensity distribution supervision (+ ILP in Table~\ref{tab:quan_res_lndb}) helped in segmenting \emph{non-solid} nodules better thus resulting in the performance improvement also for all nodules, except for the per lesion Dice scores of the 3D U-Net.
Per lesion Dice score, by definition, does not take into account FPs that are apart from GT lesions. On the other hand, it focuses on segmentation quality around GT lesions. Therefore, FNs are considered more important than FPs in calculating it. The proposed method with the 3D U-Net reduced FPs but induced FNs, which led to increased per case Dice scores but decreased per lesion Dice scores. Nevertheless, the added intensity distribution supervision on \emph{non-solid} nodules helped in segmenting them while overcoming their fuzzy appearance and underrepresentation in the dataset.

\begin{figure}[t]
	\centering
    \begin{minipage}{1\columnwidth}
        \subfloat{\includegraphics[width = 0.24\columnwidth]{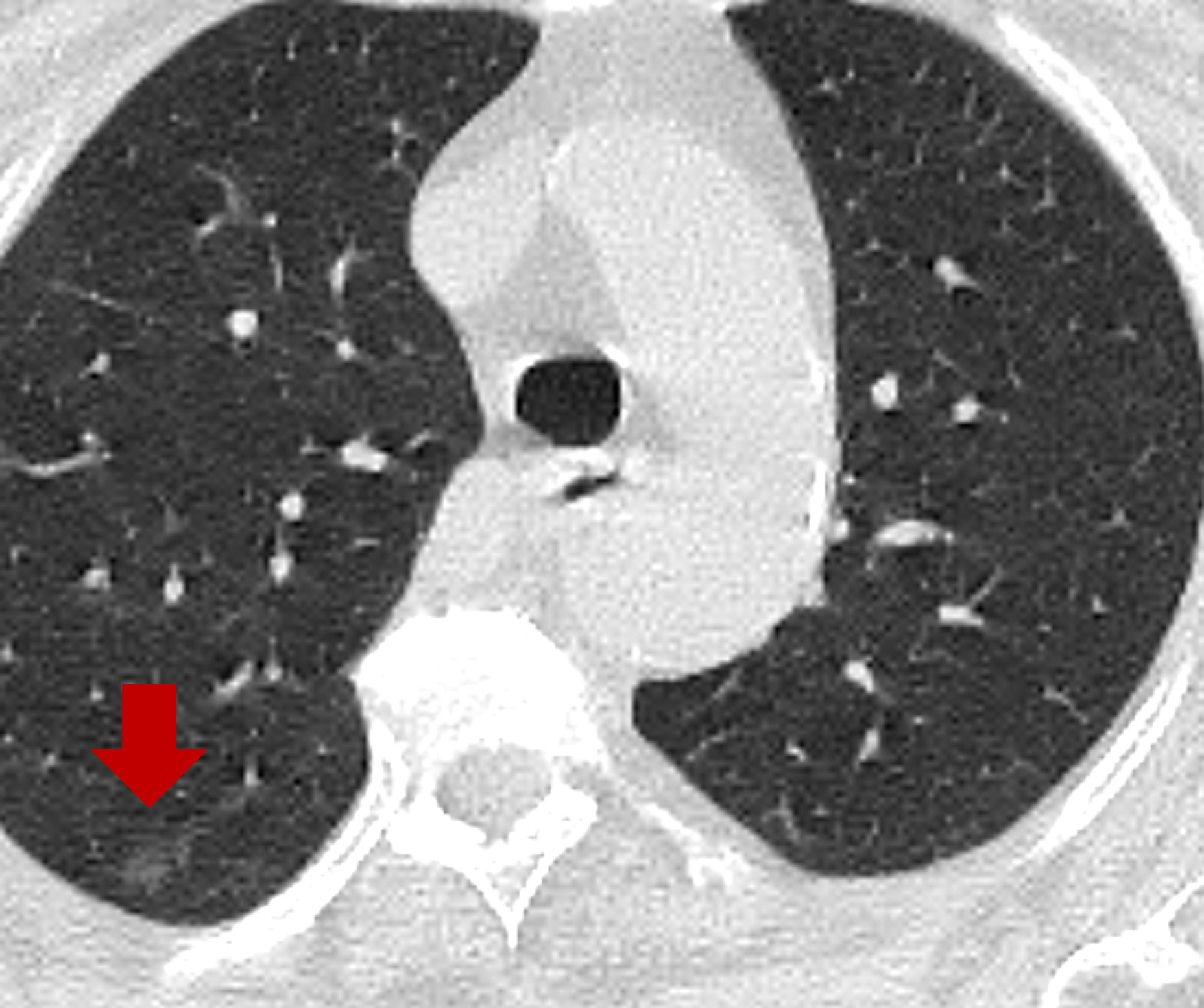}}
        \hspace{0.0001\columnwidth}
	    \subfloat{\includegraphics[width = 0.24\columnwidth]{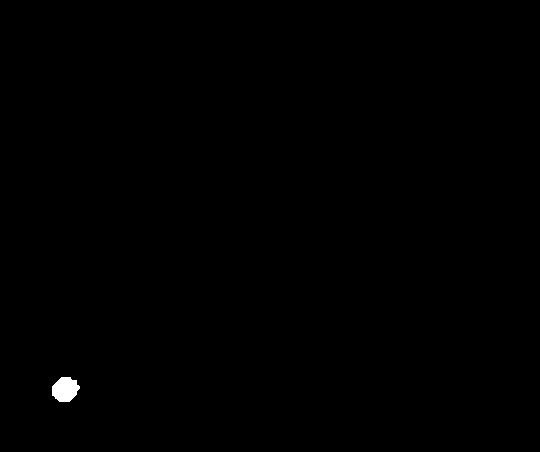}}
        \hspace{0.0001\columnwidth}
	    \subfloat{\includegraphics[width = 0.24\columnwidth]{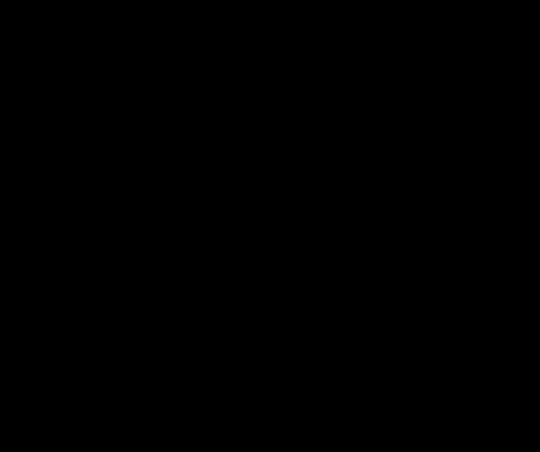}}
        \hspace{0.0001\columnwidth}
	    \subfloat{\includegraphics[width = 0.24\columnwidth]{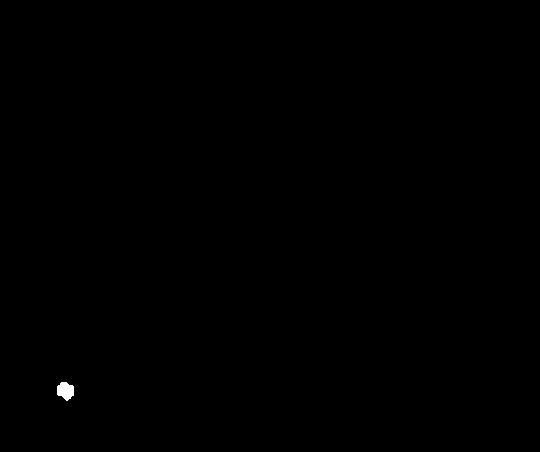}}
    \end{minipage}
    \begin{minipage}{1\columnwidth}
        \subfloat{\includegraphics[width = 0.24\columnwidth]{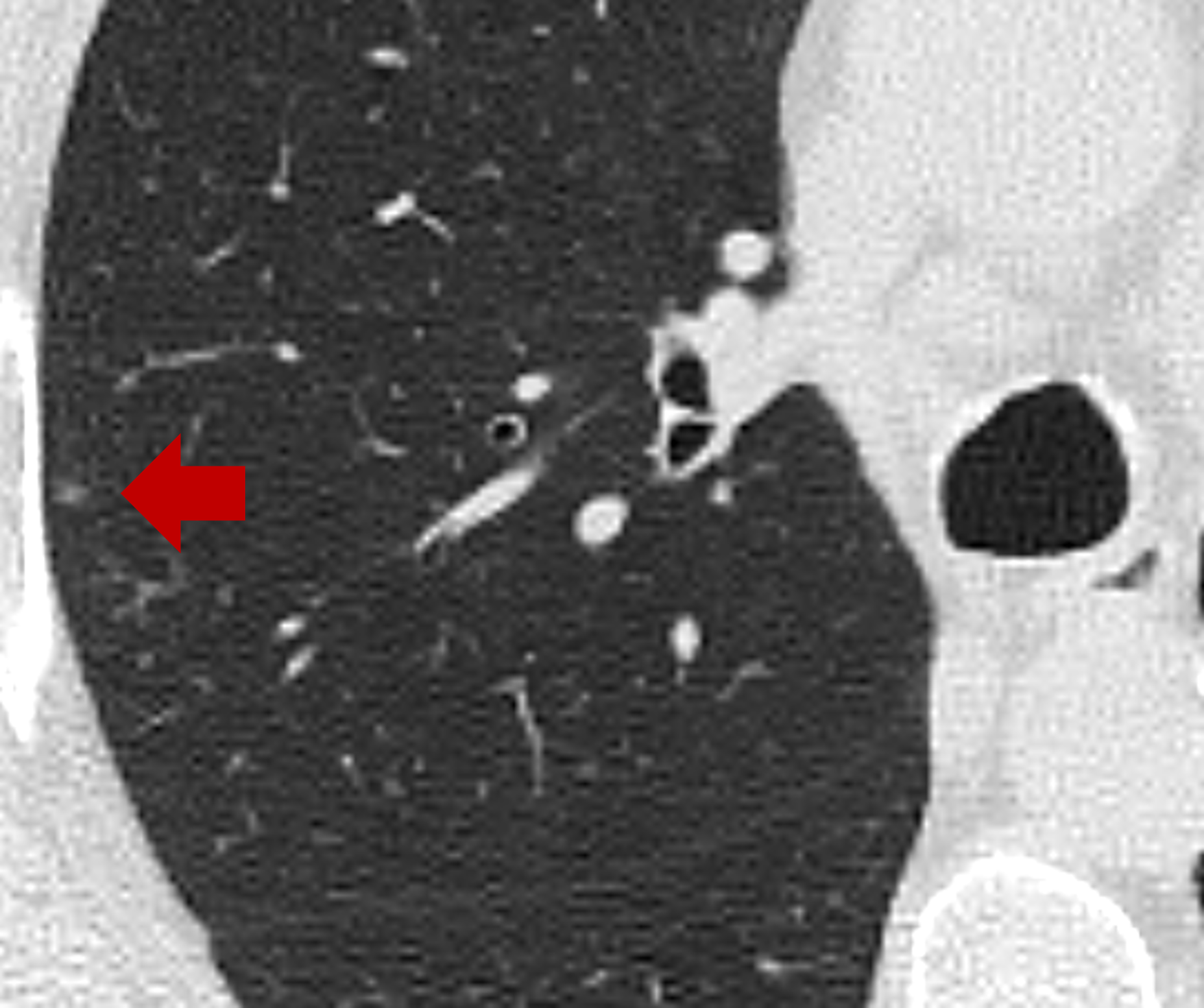}}
        \hspace{0.0001\columnwidth}
	    \subfloat{\includegraphics[width = 0.24\columnwidth]{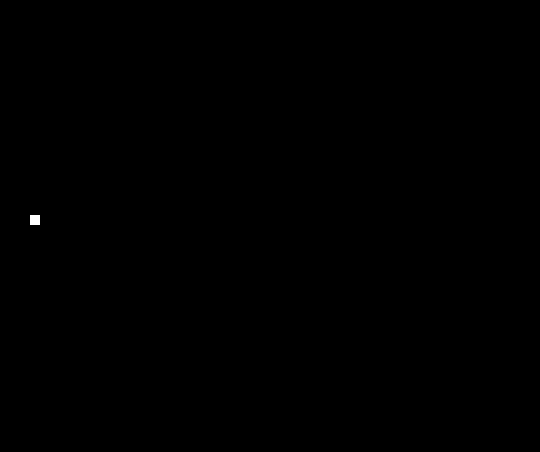}}
        \hspace{0.0001\columnwidth}
	    \subfloat{\includegraphics[width = 0.24\columnwidth]{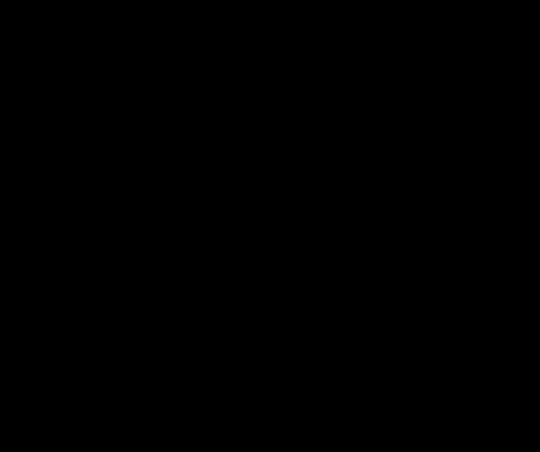}}
        \hspace{0.0001\columnwidth}
	    \subfloat{\includegraphics[width = 0.24\columnwidth]{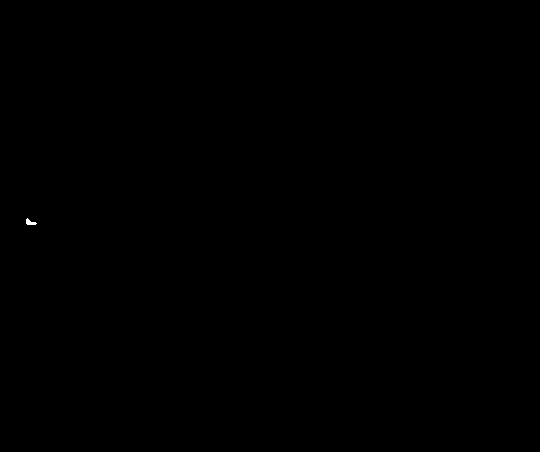}}
    \end{minipage}
    \begin{minipage}{1\columnwidth}
        \subfloat{\includegraphics[width = 0.24\columnwidth]{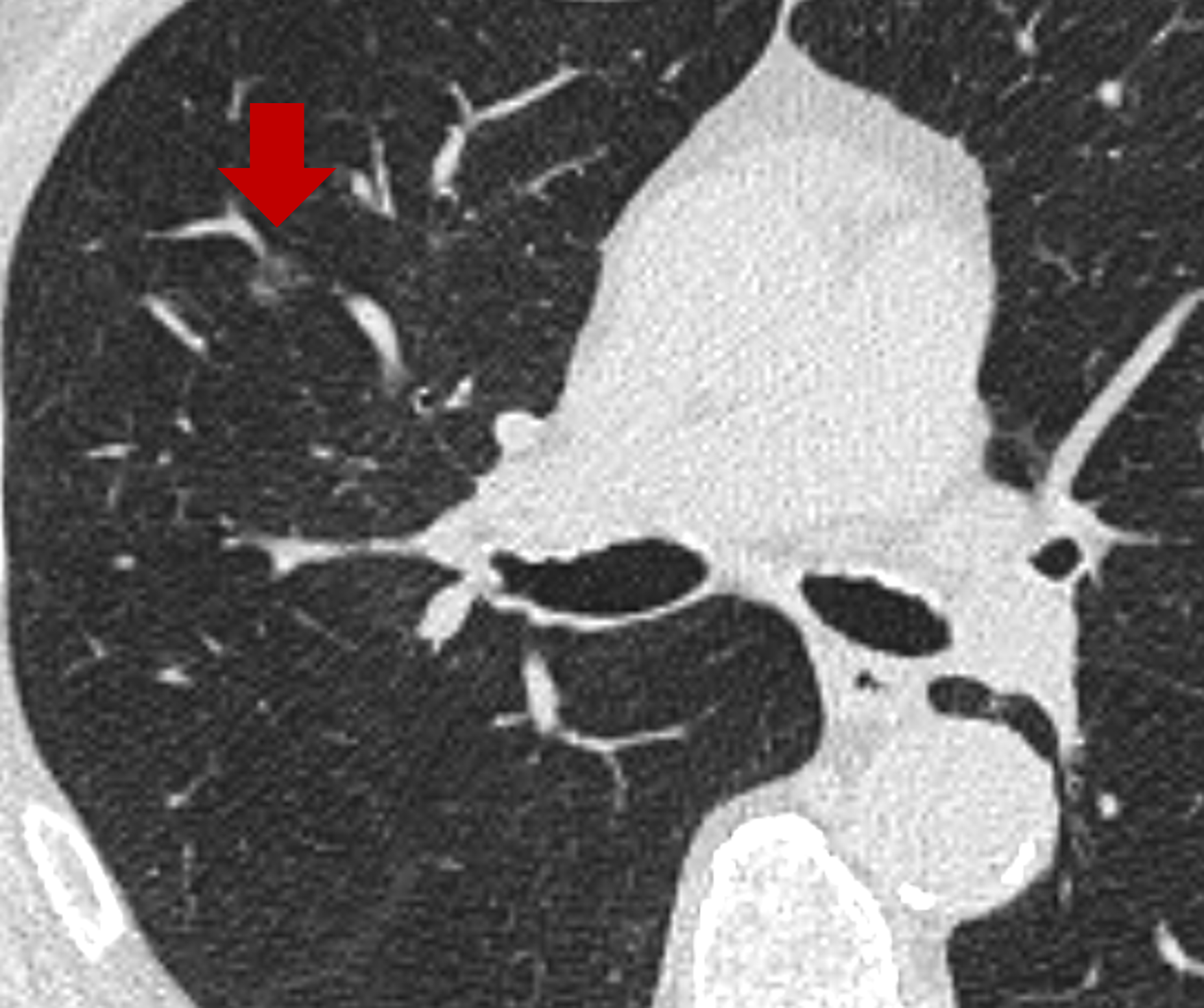}}
        \hspace{0.0001\columnwidth}
	    \subfloat{\includegraphics[width = 0.24\columnwidth]{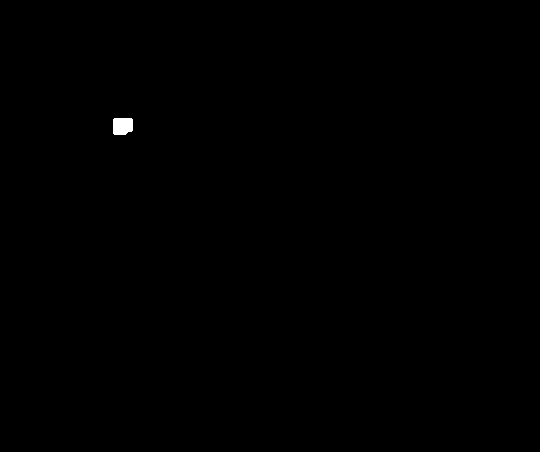}}
        \hspace{0.0001\columnwidth}
	    \subfloat{\includegraphics[width = 0.24\columnwidth]{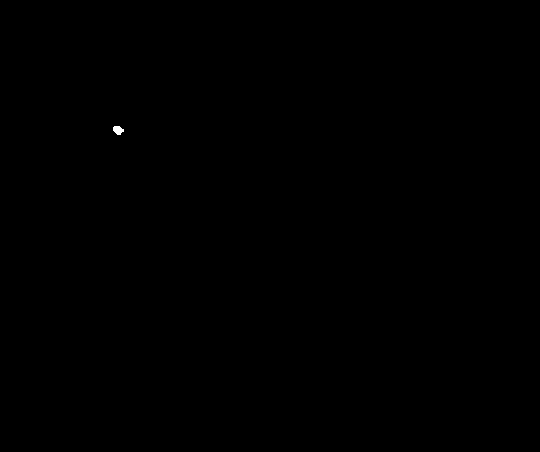}}
        \hspace{0.0001\columnwidth}
	    \subfloat{\includegraphics[width = 0.24\columnwidth]{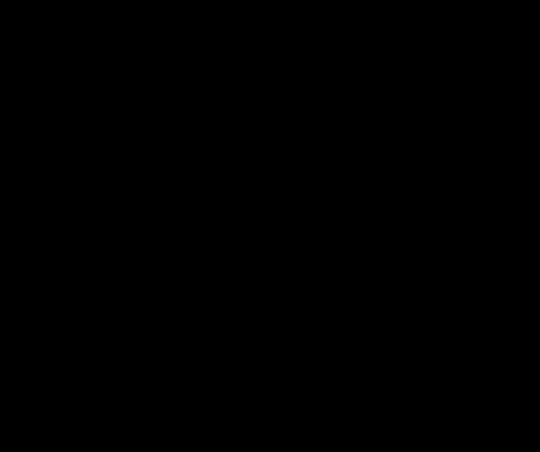}}
    \end{minipage}
    \caption{Example segmentation results on the LNDb dataset. Each row represents different cases. The columns, from left, represent input CT scan (lung nodules, which are all \emph{non-solid} in these examples, are pointed by the red arrows), corresponding GT segmentation, result of the baseline method (`nnU-Net' in Table~\ref{tab:quan_res_lndb}), and result of the proposed method (`nnU-Net + ILP' in Table~\ref{tab:quan_res_lndb}), respectively.}
	\label{fig:qual_res_lndb}
\end{figure}

\paragraph{Qualitative Results}
Figure~\ref{fig:qual_res_lndb} presents example segmentation results on the LNDb dataset. Compared to the baseline method, the proposed method segments more nodules (first and second rows).

\subsection{Lesion Detection}
\subsubsection{Quantitative Results}

\begin{table}[t]
\centering
\setlength{\tabcolsep}{3pt}
\begin{tabular}{@{}lc@{}}
\toprule
Method & Average precision (\%) \\
\midrule
Retina Net~\citep{lin17_iccv} & 64.6 \\
Retina U-Net~\citep{jaeger20} & 72.5 \\
Retina U-Net - Segm + ILP & \textbf{75.5} \\
\bottomrule
\end{tabular}
\caption{Results of different detection methods on the KiTS21 dataset. Network architecture of each method is visually explained in Figure~\ref{fig:net_det}.
`Retina U-Net - Segm + ILP' denotes the proposed method that uses the ILP supervision in place of lesion segmentation, which is used for the Retina U-Net.
}\label{tab:quan_res_kits21_det}
\end{table}

Table~\ref{tab:quan_res_kits21_det} presents quantitative results of detection methods that differ in augmenting the baseline Retina Net~\citep{lin17_iccv} on the KiTS21 dataset. The network architecture of each method is explained in Figure~\ref{fig:net_det} and the corresponding text. The Retina U-Net~\citep{jaeger20} exploiting lesion segmentation supervision, which is assumed to be available together with detection GTs, outperformed the Retina Net, as suggested in the work of \citet{jaeger20}.
When the lesion segmentation supervision was replaced with the proposed ILP supervision, it outperformed the baseline Retina Net again and further outperformed the Retina U-Net.
While the ILP function could be less precise, it can be constructed using a small number of GT lesion segmentations or can be even provided by a user as prior information. 
Also, while the Retina Net was used as the baseline here, the proposed method can be used to enhance any detectors that are based on FPNs in the same manner.

\begin{figure}[!t]
	\centering
    \begin{minipage}{1\columnwidth}
        \subfloat{\includegraphics[width = 0.33\columnwidth]{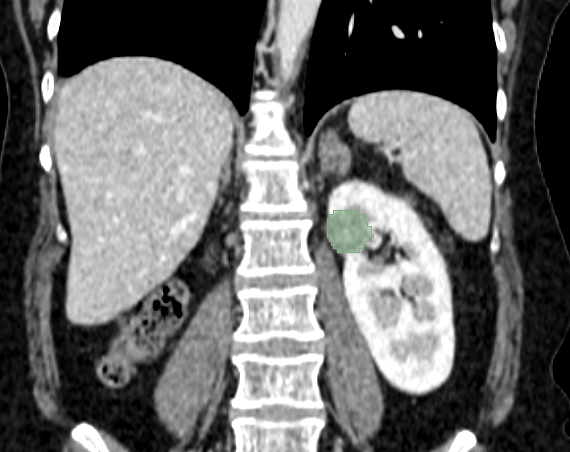}}
	    \subfloat{\includegraphics[width = 0.33\columnwidth]{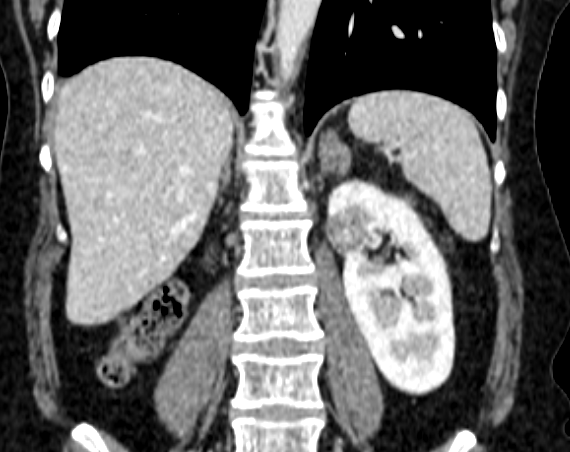}}
	    \subfloat{\includegraphics[width = 0.33\columnwidth]{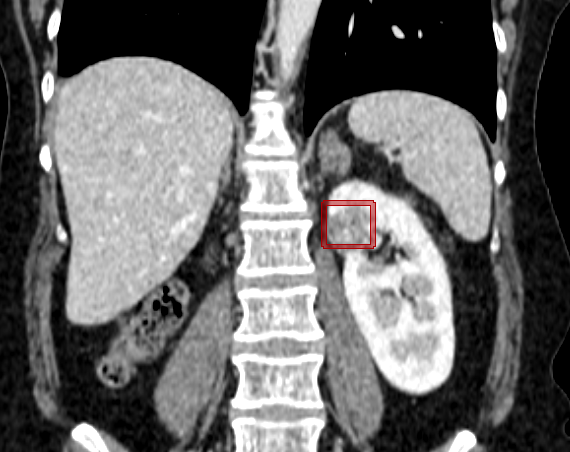}}
        \vspace{-0.35cm}
    \end{minipage}
    \begin{minipage}{1\columnwidth}
        \subfloat{\includegraphics[width = 0.33\columnwidth]{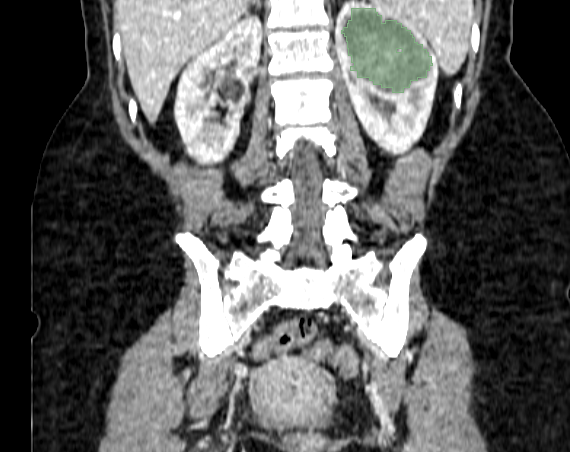}}
	    \subfloat{\includegraphics[width = 0.33\columnwidth]{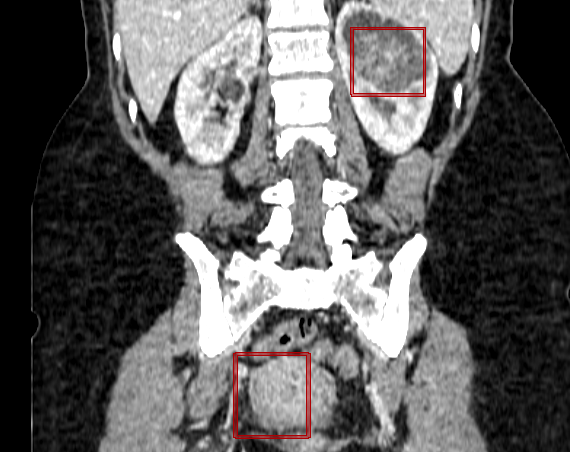}}
	    \subfloat{\includegraphics[width = 0.33\columnwidth]{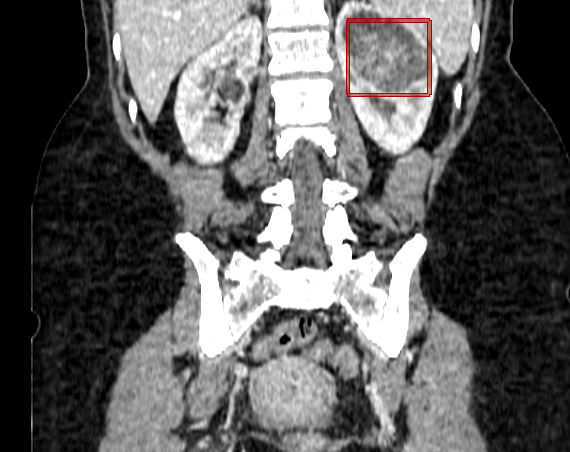}}
        \vspace{-0.35cm}
    \end{minipage}
    \begin{minipage}{1\columnwidth}
        \subfloat{\includegraphics[width = 0.33\columnwidth]{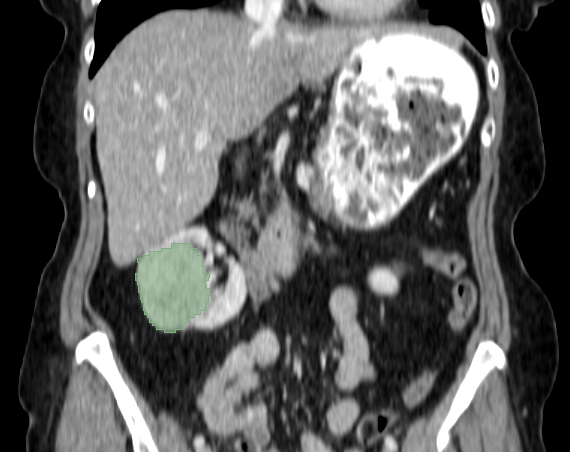}}
	    \subfloat{\includegraphics[width = 0.33\columnwidth]{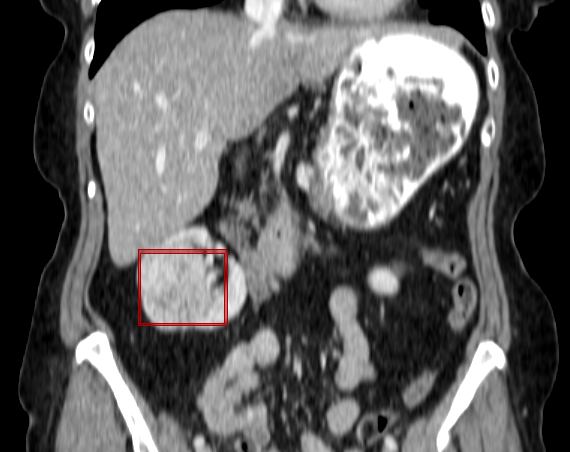}}
	    \subfloat{\includegraphics[width = 0.33\columnwidth]{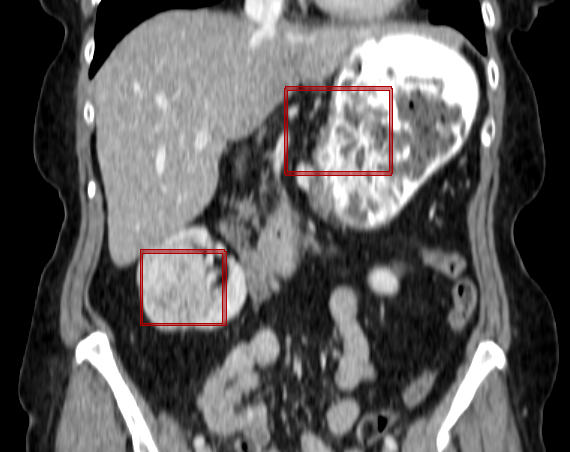}}
        \vspace{-0.35cm}
    \end{minipage}
    \begin{minipage}{1\columnwidth}
        \subfloat{\includegraphics[width = 0.33\columnwidth]{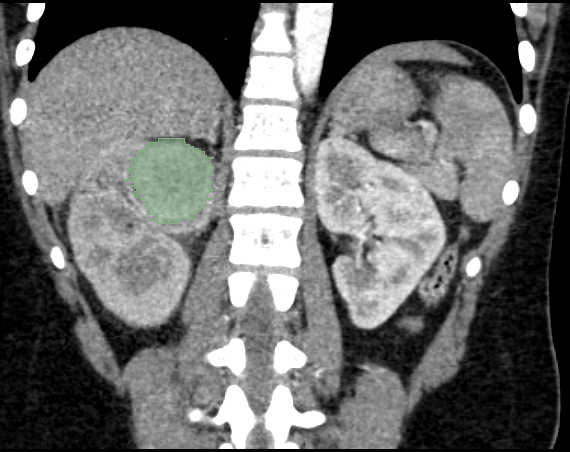}}
	    \subfloat{\includegraphics[width = 0.33\columnwidth]{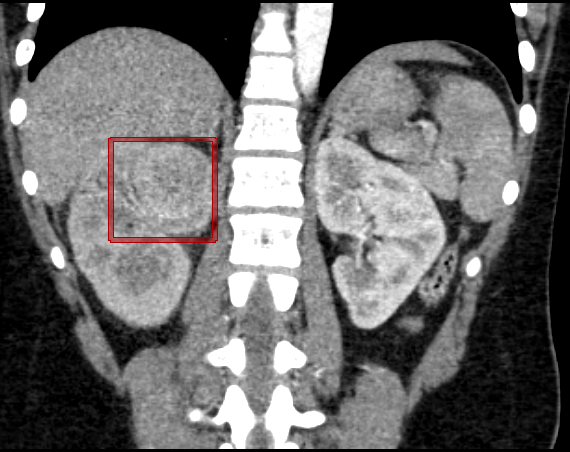}}
	    \subfloat{\includegraphics[width = 0.33\columnwidth]{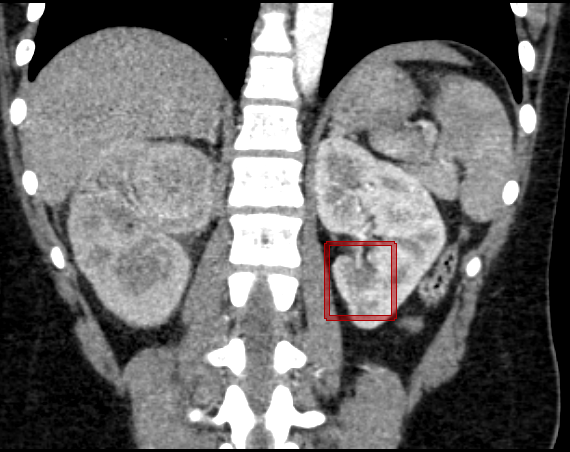}}
    \end{minipage}
    \caption{Example detection results on the KiTS21 dataset. Each row represents different cases. The columns, from left, represent input CT scan (available kidney tumor segmentation is overlaid to locate the tumors), result of the baseline method (`Retina Net' in Table~\ref{tab:quan_res_kits21_det}), and result of the proposed method (`Retina U-Net - Segm + ILP' in Table~\ref{tab:quan_res_kits21_det}), respectively.}
	\label{fig:qual_res_kits21_det}
\end{figure}

\subsubsection{Qualitative Results}

Figure~\ref{fig:qual_res_kits21_det} shows example detection results on the KiTS21 dataset.
The incorporated intensity distribution information helped in locating a tumor (first row) and eliminating a false positive (second row).
The bottom two rows represent failure cases. In the third row, a false positive was detected by the proposed method on the heterogeneous stomach, which resembles a kidney with a tumor in appearance. In the last row, the tumor that has similar intensity values as the rest of the kidney was missed by the proposed method.

\section{Discussion}
We have presented a method to incorporate the intensity information of a target lesion on CT scans in training segmentation and detection networks.
An ILP function constructed from an intensity histogram of a target lesion is used to effectively locate regions where the lesions are possibly situated. The ILP map of each input CT scan is provided as additional supervision for network training.
It aims to inform the network about our region of interest, which could contain target lesions, especially in terms of intensity values. It requires no additional labeling effort.

The method has been applied to improve the segmentation of three different lesion types, namely, small bowel carcinoid tumors, kidney tumors, and lung nodules. The effectiveness of the proposed method on a detection task has been also investigated for kidney tumors.
Our findings from the experiments are: 1) The proposed method of using the ILP as additional supervision performs better than other usages of it, such as for post-processing and as an additional input channel (Table~\ref{tab:quan_res_sbct}).
2) Having a precise and generalizable intensity distribution is important for the success of the method (Table~\ref{tab:quan_res_sbct}).
3) It can be effectively used with the nnU-Net for more optimized performance (Table~\ref{tab:quan_res_kits21_segm}). 4) It performs favorably against a method that exploits another supervision such as organ segmentation (Table~\ref{tab:quan_res_kits21_segm}). 5) Consistent performance gains can be expected over varying training set sizes (Figure~\ref{fig:quan_res_kits21_segm_by_train_size}). 6) It can be considered to boost the performance of an underrepresented lesion type (Table~\ref{tab:quan_res_lndb}).
7) It can be used to enhance a detector such as the Retina Net~\citep{lin17_iccv} (Table~\ref{tab:quan_res_kits21_det}).

Carcinoid tumors in our SBCT dataset are small (Figure~\ref{fig:qual_res_sbct}). Lung nodules in the LNDb dataset are also small (mostly less than a centimeter~\citep{pedrosa19}) as exemplified in Figure~\ref{fig:qual_res_lndb}. Even small numbers of false positive and false negative voxels have a big impact on the Dice score of small lesions. Nevertheless, the proposed method showed clear improvements compared to the baseline.
We also note that we segmented nodules from an entire CT scan, whereas the segmentation is conducted when nodule centroids are given for each scan in the LNDb challenge.
Our task is more challenging, which makes achieving high Dice scores difficult again.

For the KiTS21 dataset, we incorporated the proposed intensity distribution supervision also in training the challenge winning method~\citep{zhao21}. Although the efficacy of the proposed method was verified, our result is not directly comparable with theirs since they used more training images (240 vs. 210). While they divided the dataset into only training and validation sets (a separate test set available for the challenge period), we divided it into training/validation/test sets to enable a strict evaluation within the available data.
Also, we did not use their postprocessing method of counting the number of voxels for each connected component and thresholding them based on their sizes, since that heuristics is not always relevant.

In terms of network training, typical segmentation and detection losses are used together with the ILP loss for the segmentation and detection tasks, respectively (Eq.~\ref{eq:loss_net_segm} and Eq.~\ref{eq:loss_net_det}).
The proposed method provides an \emph{additional opportunity} to consider the intensity information of the target lesion in an explicit way while retaining learning about other aspects by the typical loss terms.
A lesion that is not distinct by the ILP model still can be identified by the other aspects.
For example, in the third row of Figure~\ref{fig:qual_res_kits21_det}, the tumor that is not distinct from the kidney by intensity values was still detected by the proposed method.

In this work, for each target lesion, the experiments have been conducted on a single dataset acquired using a particular imaging protocol. The proposed method would be less applicable across datasets that were acquired using different imaging protocols since the intensity distribution of the target lesion can be diffused and incoherent. Also, we took a relatively simple implementation for incorporating the intensity information of the target lesion in the network training. For the same objective, a better approach can be explored.

In future work, we plan to study the effect of incorporating unsupervised images into training since the proposed intensity distribution supervision enables training on them. The proposed method can be further applied to different target lesions.

\section*{Data Availability}
The code is available at \href{https://github.com/rsummers11/CADLab/tree/master/intensity_distribution_supervision}{https://github.com/rsummers11/CADLab/tree\\/master/intensity\_distribution\_supervision}

\section*{Acknowledgment}
This research was supported by the Intramural Research Program of the National Institutes of Health, Clinical Center. The research used the high-performance computing facilities of the NIH Biowulf cluster.

\section*{Conflicts of Interest}
Potential financial interest: Author Ronald M. Summers receives royalties from iCAD, Philips, Scan Med, PingAn, and Translation Holdings and has received research support from Ping An (CRADA).


\begin{thebibliography}{37}
	\expandafter\ifx\csname natexlab\endcsname\relax\def\natexlab#1{#1}\fi
	\providecommand{\url}[1]{\texttt{#1}}
	\providecommand{\href}[2]{#2}
	\providecommand{\path}[1]{#1}
	\providecommand{\DOIprefix}{doi:}
	\providecommand{\ArXivprefix}{arXiv:}
	\providecommand{\URLprefix}{URL: }
	\providecommand{\Pubmedprefix}{pmid:}
	\providecommand{\doi}[1]{\href{http://dx.doi.org/#1}{\path{#1}}}
	\providecommand{\Pubmed}[1]{\href{pmid:#1}{\path{#1}}}
	\providecommand{\bibinfo}[2]{#2}
	\ifx\xfnm\relax \def\xfnm[#1]{\unskip,\space#1}\fi
	\bibitem[{{American Cancer Society}(2022)}]{acs_lung_cancer}
	\bibinfo{author}{{American Cancer Society}}, \bibinfo{year}{2022}.
	\newblock \bibinfo{title}{Key statistics for lung cancer}.
	\newblock \URLprefix
	\url{https://www.cancer.org/content/dam/CRC/PDF/Public/8703.00.pdf}.
	\bibitem[{Ayalew et~al.(2021)Ayalew, Fante and Mohammed}]{ayalew21}
	\bibinfo{author}{Ayalew, Y.A.}, \bibinfo{author}{Fante, K.A.},
	\bibinfo{author}{Mohammed, M.A.}, \bibinfo{year}{2021}.
	\newblock \bibinfo{title}{Modified u-net for liver cancer segmentation from
		computed tomography images with a new class balancing method}.
	\newblock \bibinfo{journal}{BMC Biomedical Engineering} \bibinfo{volume}{3},
	\bibinfo{pages}{1--13}.
	\bibitem[{Black et~al.(1998)Black, Sapiro, Marimont and Heeger}]{black98}
	\bibinfo{author}{Black, M.}, \bibinfo{author}{Sapiro, G.},
	\bibinfo{author}{Marimont, D.}, \bibinfo{author}{Heeger, D.},
	\bibinfo{year}{1998}.
	\newblock \bibinfo{title}{Robust anisotropic diffusion}.
	\newblock \bibinfo{journal}{IEEE Transactions on Image Processing}
	\bibinfo{volume}{7}, \bibinfo{pages}{421--432}.
	\newblock \DOIprefix\doi{10.1109/83.661192}.
	\bibitem[{Buzug(2011)}]{buzug11}
	\bibinfo{author}{Buzug, T.M.}, \bibinfo{year}{2011}.
	\newblock \bibinfo{title}{Computed tomography}, in:
	\bibinfo{booktitle}{Springer handbook of medical technology}.
	\bibinfo{publisher}{Springer}, pp. \bibinfo{pages}{311--342}.
	\bibitem[{Cai et~al.(2021)Cai, Tang, Yan, Harrison, Xiao, Lin and Lu}]{cai21}
	\bibinfo{author}{Cai, J.}, \bibinfo{author}{Tang, Y.}, \bibinfo{author}{Yan,
		K.}, \bibinfo{author}{Harrison, A.P.}, \bibinfo{author}{Xiao, J.},
	\bibinfo{author}{Lin, G.}, \bibinfo{author}{Lu, L.}, \bibinfo{year}{2021}.
	\newblock \bibinfo{title}{Deep lesion tracker: Monitoring lesions in 4d
		longitudinal imaging studies}, in: \bibinfo{booktitle}{2021 IEEE/CVF
		Conference on Computer Vision and Pattern Recognition (CVPR)}, pp.
	\bibinfo{pages}{15154--15164}.
	\newblock \DOIprefix\doi{10.1109/CVPR46437.2021.01491}.
	\bibitem[{Cancer.Net(2022)}]{cancernet_kidney_cancer}
	\bibinfo{author}{Cancer.Net}, \bibinfo{year}{2022}.
	\newblock \bibinfo{title}{Kidney cancer: Statistics}.
	\newblock \URLprefix
	\url{https://www.cancer.net/cancer-types/kidney-cancer/statistics}.
	\bibitem[{{\c{C}}i{\c{c}}ek et~al.(2016){\c{C}}i{\c{c}}ek, Abdulkadir,
		Lienkamp, Brox and Ronneberger}]{cicek16}
	\bibinfo{author}{{\c{C}}i{\c{c}}ek, {\"O}.}, \bibinfo{author}{Abdulkadir, A.},
	\bibinfo{author}{Lienkamp, S.S.}, \bibinfo{author}{Brox, T.},
	\bibinfo{author}{Ronneberger, O.}, \bibinfo{year}{2016}.
	\newblock \bibinfo{title}{3d u-net: Learning dense volumetric segmentation from
		sparse annotation}, in: \bibinfo{editor}{Ourselin, S.},
	\bibinfo{editor}{Joskowicz, L.}, \bibinfo{editor}{Sabuncu, M.R.},
	\bibinfo{editor}{Unal, G.}, \bibinfo{editor}{Wells, W.} (Eds.),
	\bibinfo{booktitle}{Medical Image Computing and Computer-Assisted
		Intervention -- MICCAI 2016}, \bibinfo{publisher}{Springer International
		Publishing}, \bibinfo{address}{Cham}. pp. \bibinfo{pages}{424--432}.
	\bibitem[{Diederich(2009)}]{diederich09}
	\bibinfo{author}{Diederich, S.}, \bibinfo{year}{2009}.
	\newblock \bibinfo{title}{Pulmonary nodules: do we need a separate algorithm
		for non-solid lesions?}
	\newblock \bibinfo{journal}{Cancer Imaging} \bibinfo{volume}{9},
	\bibinfo{pages}{S126}.
	\bibitem[{Eisenhauer et~al.(2009)Eisenhauer, Therasse, Bogaerts, Schwartz,
		Sargent, Ford, Dancey, Arbuck, Gwyther, Mooney et~al.}]{eisenhauer09}
	\bibinfo{author}{Eisenhauer, E.A.}, \bibinfo{author}{Therasse, P.},
	\bibinfo{author}{Bogaerts, J.}, \bibinfo{author}{Schwartz, L.H.},
	\bibinfo{author}{Sargent, D.}, \bibinfo{author}{Ford, R.},
	\bibinfo{author}{Dancey, J.}, \bibinfo{author}{Arbuck, S.},
	\bibinfo{author}{Gwyther, S.}, \bibinfo{author}{Mooney, M.}, et~al.,
	\bibinfo{year}{2009}.
	\newblock \bibinfo{title}{New response evaluation criteria in solid tumours:
		revised recist guideline (version 1.1)}.
	\newblock \bibinfo{journal}{European journal of cancer} \bibinfo{volume}{45},
	\bibinfo{pages}{228--247}.
	\bibitem[{Fedorov et~al.(2012)Fedorov, Beichel, Kalpathy-Cramer, Finet,
		Fillion-Robin, Pujol, Bauer, Jennings, Fennessy, Sonka, Buatti, Aylward,
		Miller, Pieper and Kikinis}]{fedorov12}
	\bibinfo{author}{Fedorov, A.}, \bibinfo{author}{Beichel, R.},
	\bibinfo{author}{Kalpathy-Cramer, J.}, \bibinfo{author}{Finet, J.},
	\bibinfo{author}{Fillion-Robin, J.C.}, \bibinfo{author}{Pujol, S.},
	\bibinfo{author}{Bauer, C.}, \bibinfo{author}{Jennings, D.},
	\bibinfo{author}{Fennessy, F.}, \bibinfo{author}{Sonka, M.},
	\bibinfo{author}{Buatti, J.}, \bibinfo{author}{Aylward, S.},
	\bibinfo{author}{Miller, J.V.}, \bibinfo{author}{Pieper, S.},
	\bibinfo{author}{Kikinis, R.}, \bibinfo{year}{2012}.
	\newblock \bibinfo{title}{3d slicer as an image computing platform for the
		quantitative imaging network}.
	\newblock \bibinfo{journal}{Magnetic Resonance Imaging} \bibinfo{volume}{30},
	\bibinfo{pages}{1323 -- 1341}.
	\newblock \DOIprefix\doi{https://doi.org/10.1016/j.mri.2012.05.001}.
	\bibinfo{note}{quantitative Imaging in Cancer}.
	\bibitem[{He et~al.(2016)He, Zhang, Ren and Sun}]{he16}
	\bibinfo{author}{He, K.}, \bibinfo{author}{Zhang, X.}, \bibinfo{author}{Ren,
		S.}, \bibinfo{author}{Sun, J.}, \bibinfo{year}{2016}.
	\newblock \bibinfo{title}{Deep residual learning for image recognition}, in:
	\bibinfo{booktitle}{2016 IEEE Conference on Computer Vision and Pattern
		Recognition (CVPR)}, pp. \bibinfo{pages}{770--778}.
	\newblock \DOIprefix\doi{10.1109/CVPR.2016.90}.
	\bibitem[{Heller et~al.(2021)Heller, Isensee, Maier-Hein, Hou, Xie, Li, Nan,
		Mu, Lin, Han, Yao, Gao, Zhang, Wang, Hou, Yang, Xiong, Tian, Zhong, Ma,
		Rickman, Dean, Stai, Tejpaul, Oestreich, Blake, Kaluzniak, Raza, Rosenberg,
		Moore, Walczak, Rengel, Edgerton, Vasdev, Peterson, McSweeney, Peterson,
		Kalapara, Sathianathen, Papanikolopoulos and Weight}]{heller21}
	\bibinfo{author}{Heller, N.}, \bibinfo{author}{Isensee, F.},
	\bibinfo{author}{Maier-Hein, K.H.}, \bibinfo{author}{Hou, X.},
	\bibinfo{author}{Xie, C.}, \bibinfo{author}{Li, F.}, \bibinfo{author}{Nan,
		Y.}, \bibinfo{author}{Mu, G.}, \bibinfo{author}{Lin, Z.},
	\bibinfo{author}{Han, M.}, \bibinfo{author}{Yao, G.}, \bibinfo{author}{Gao,
		Y.}, \bibinfo{author}{Zhang, Y.}, \bibinfo{author}{Wang, Y.},
	\bibinfo{author}{Hou, F.}, \bibinfo{author}{Yang, J.},
	\bibinfo{author}{Xiong, G.}, \bibinfo{author}{Tian, J.},
	\bibinfo{author}{Zhong, C.}, \bibinfo{author}{Ma, J.},
	\bibinfo{author}{Rickman, J.}, \bibinfo{author}{Dean, J.},
	\bibinfo{author}{Stai, B.}, \bibinfo{author}{Tejpaul, R.},
	\bibinfo{author}{Oestreich, M.}, \bibinfo{author}{Blake, P.},
	\bibinfo{author}{Kaluzniak, H.}, \bibinfo{author}{Raza, S.},
	\bibinfo{author}{Rosenberg, J.}, \bibinfo{author}{Moore, K.},
	\bibinfo{author}{Walczak, E.}, \bibinfo{author}{Rengel, Z.},
	\bibinfo{author}{Edgerton, Z.}, \bibinfo{author}{Vasdev, R.},
	\bibinfo{author}{Peterson, M.}, \bibinfo{author}{McSweeney, S.},
	\bibinfo{author}{Peterson, S.}, \bibinfo{author}{Kalapara, A.},
	\bibinfo{author}{Sathianathen, N.}, \bibinfo{author}{Papanikolopoulos, N.},
	\bibinfo{author}{Weight, C.}, \bibinfo{year}{2021}.
	\newblock \bibinfo{title}{The state of the art in kidney and kidney tumor
		segmentation in contrast-enhanced ct imaging: Results of the kits19
		challenge}.
	\newblock \bibinfo{journal}{Medical Image Analysis} \bibinfo{volume}{67},
	\bibinfo{pages}{101821}.
	\newblock \DOIprefix\doi{https://doi.org/10.1016/j.media.2020.101821}.
	\bibitem[{Hughes et~al.(2016)Hughes, Azoury, Assadipour, Straughan, Trivedi,
		Lim, Joy, Voellinger, Tang, Venkatesan et~al.}]{hughes16}
	\bibinfo{author}{Hughes, M.S.}, \bibinfo{author}{Azoury, S.C.},
	\bibinfo{author}{Assadipour, Y.}, \bibinfo{author}{Straughan, D.M.},
	\bibinfo{author}{Trivedi, A.N.}, \bibinfo{author}{Lim, R.M.},
	\bibinfo{author}{Joy, G.}, \bibinfo{author}{Voellinger, M.T.},
	\bibinfo{author}{Tang, D.M.}, \bibinfo{author}{Venkatesan, A.M.}, et~al.,
	\bibinfo{year}{2016}.
	\newblock \bibinfo{title}{Prospective evaluation and treatment of familial
		carcinoid small intestine neuroendocrine tumors (si-nets)}.
	\newblock \bibinfo{journal}{Surgery} \bibinfo{volume}{159},
	\bibinfo{pages}{350--357}.
	\bibitem[{Isensee et~al.(2021)Isensee, Jaeger, Kohl, Petersen and
		Maier-Hein}]{isensee21}
	\bibinfo{author}{Isensee, F.}, \bibinfo{author}{Jaeger, P.F.},
	\bibinfo{author}{Kohl, S.A.}, \bibinfo{author}{Petersen, J.},
	\bibinfo{author}{Maier-Hein, K.H.}, \bibinfo{year}{2021}.
	\newblock \bibinfo{title}{nnu-net: a self-configuring method for deep
		learning-based biomedical image segmentation}.
	\newblock \bibinfo{journal}{Nature methods} \bibinfo{volume}{18},
	\bibinfo{pages}{203--211}.
	\bibitem[{Jaeger et~al.(2020)Jaeger, Kohl, Bickelhaupt, Isensee, Kuder,
		Schlemmer and Maier-Hein}]{jaeger20}
	\bibinfo{author}{Jaeger, P.F.}, \bibinfo{author}{Kohl, S.A.A.},
	\bibinfo{author}{Bickelhaupt, S.}, \bibinfo{author}{Isensee, F.},
	\bibinfo{author}{Kuder, T.A.}, \bibinfo{author}{Schlemmer, H.P.},
	\bibinfo{author}{Maier-Hein, K.H.}, \bibinfo{year}{2020}.
	\newblock \bibinfo{title}{{Retina U-Net: Embarrassingly Simple Exploitation of
			Segmentation Supervision for Medical Object Detection}}, in:
	\bibinfo{editor}{Dalca, A.V.}, \bibinfo{editor}{McDermott, M.B.},
	\bibinfo{editor}{Alsentzer, E.}, \bibinfo{editor}{Finlayson, S.G.},
	\bibinfo{editor}{Oberst, M.}, \bibinfo{editor}{Falck, F.},
	\bibinfo{editor}{Beaulieu-Jones, B.} (Eds.), \bibinfo{booktitle}{Proceedings
		of the Machine Learning for Health NeurIPS Workshop},
	\bibinfo{publisher}{PMLR}. pp. \bibinfo{pages}{171--183}.
	\newblock \URLprefix \url{https://proceedings.mlr.press/v116/jaeger20a.html}.
	\bibitem[{Jasti and Carucci(2020)}]{jasti20}
	\bibinfo{author}{Jasti, R.}, \bibinfo{author}{Carucci, L.R.},
	\bibinfo{year}{2020}.
	\newblock \bibinfo{title}{Small bowel neoplasms: A pictorial review}.
	\newblock \bibinfo{journal}{RadioGraphics} \bibinfo{volume}{40},
	\bibinfo{pages}{1020--1038}.
	\newblock \URLprefix \url{https://doi.org/10.1148/rg.2020200011},
	\DOIprefix\doi{10.1148/rg.2020200011},
	\href{http://arxiv.org/abs/https://doi.org/10.1148/rg.2020200011}{{\tt
			arXiv:https://doi.org/10.1148/rg.2020200011}}. \bibinfo{note}{pMID:
		32559148}.
	\bibitem[{Kamble et~al.(2020)Kamble, Sahu and Doriya}]{kamble20}
	\bibinfo{author}{Kamble, B.}, \bibinfo{author}{Sahu, S.P.},
	\bibinfo{author}{Doriya, R.}, \bibinfo{year}{2020}.
	\newblock \bibinfo{title}{A review on lung and nodule segmentation techniques},
	in: \bibinfo{editor}{Kolhe, M.L.}, \bibinfo{editor}{Tiwari, S.},
	\bibinfo{editor}{Trivedi, M.C.}, \bibinfo{editor}{Mishra, K.K.} (Eds.),
	\bibinfo{booktitle}{Advances in Data and Information Sciences},
	\bibinfo{publisher}{Springer Singapore}, \bibinfo{address}{Singapore}. pp.
	\bibinfo{pages}{555--565}.
	\bibitem[{Kingma and Ba(2015)}]{kingma15}
	\bibinfo{author}{Kingma, D.P.}, \bibinfo{author}{Ba, J.}, \bibinfo{year}{2015}.
	\newblock \bibinfo{title}{Adam: {A} method for stochastic optimization}, in:
	\bibinfo{editor}{Bengio, Y.}, \bibinfo{editor}{LeCun, Y.} (Eds.),
	\bibinfo{booktitle}{3rd International Conference on Learning Representations,
		{ICLR} 2015, San Diego, CA, USA, May 7-9, 2015, Conference Track
		Proceedings}.
	\newblock \URLprefix \url{http://arxiv.org/abs/1412.6980}.
	\bibitem[{Lin et~al.(2017a)Lin, Dollar, Girshick, He, Hariharan and
		Belongie}]{lin17_cvpr}
	\bibinfo{author}{Lin, T.Y.}, \bibinfo{author}{Dollar, P.},
	\bibinfo{author}{Girshick, R.}, \bibinfo{author}{He, K.},
	\bibinfo{author}{Hariharan, B.}, \bibinfo{author}{Belongie, S.},
	\bibinfo{year}{2017}a.
	\newblock \bibinfo{title}{Feature pyramid networks for object detection}, in:
	\bibinfo{booktitle}{Proceedings of the IEEE Conference on Computer Vision and
		Pattern Recognition (CVPR)}.
	\bibitem[{Lin et~al.(2017b)Lin, Goyal, Girshick, He and Dollar}]{lin17_iccv}
	\bibinfo{author}{Lin, T.Y.}, \bibinfo{author}{Goyal, P.},
	\bibinfo{author}{Girshick, R.}, \bibinfo{author}{He, K.},
	\bibinfo{author}{Dollar, P.}, \bibinfo{year}{2017}b.
	\newblock \bibinfo{title}{Focal loss for dense object detection}, in:
	\bibinfo{booktitle}{Proceedings of the IEEE International Conference on
		Computer Vision (ICCV)}.
	\bibitem[{Lin et~al.(2016)Lin, Hsu, Ko, Chu, Chou, Chang and Chang}]{lin16}
	\bibinfo{author}{Lin, Y.P.}, \bibinfo{author}{Hsu, H.H.}, \bibinfo{author}{Ko,
		K.H.}, \bibinfo{author}{Chu, C.M.}, \bibinfo{author}{Chou, Y.C.},
	\bibinfo{author}{Chang, W.C.}, \bibinfo{author}{Chang, T.H.},
	\bibinfo{year}{2016}.
	\newblock \bibinfo{title}{Differentiation of malignant and benign incidental
		breast lesions detected by chest multidetector-row computed tomography: Added
		value of quantitative enhancement analysis}.
	\newblock \bibinfo{journal}{PLOS ONE} \bibinfo{volume}{11},
	\bibinfo{pages}{1--11}.
	\newblock \URLprefix \url{https://doi.org/10.1371/journal.pone.0154569},
	\DOIprefix\doi{10.1371/journal.pone.0154569}.
	\bibitem[{Liu et~al.(2021)Liu, Tsui and Mandal}]{liu21}
	\bibinfo{author}{Liu, L.}, \bibinfo{author}{Tsui, Y.Y.},
	\bibinfo{author}{Mandal, M.}, \bibinfo{year}{2021}.
	\newblock \bibinfo{title}{Skin lesion segmentation using deep learning with
		auxiliary task}.
	\newblock \bibinfo{journal}{Journal of Imaging} \bibinfo{volume}{7}.
	\newblock \URLprefix \url{https://www.mdpi.com/2313-433X/7/4/67},
	\DOIprefix\doi{10.3390/jimaging7040067}.
	\bibitem[{Loshchilov and Hutter(2019)}]{loshchilov19}
	\bibinfo{author}{Loshchilov, I.}, \bibinfo{author}{Hutter, F.},
	\bibinfo{year}{2019}.
	\newblock \bibinfo{title}{Decoupled weight decay regularization}, in:
	\bibinfo{booktitle}{International Conference on Learning Representations}.
	\newblock \URLprefix \url{https://openreview.net/forum?id=Bkg6RiCqY7}.
	\bibitem[{Ma et~al.(2020)Ma, Wei, Zhang, Wang, Lv, Zhu, Gaoxiang, Liu, Peng,
		Wang, Wang and Chen}]{ma20}
	\bibinfo{author}{Ma, J.}, \bibinfo{author}{Wei, Z.}, \bibinfo{author}{Zhang,
		Y.}, \bibinfo{author}{Wang, Y.}, \bibinfo{author}{Lv, R.},
	\bibinfo{author}{Zhu, C.}, \bibinfo{author}{Gaoxiang, C.},
	\bibinfo{author}{Liu, J.}, \bibinfo{author}{Peng, C.}, \bibinfo{author}{Wang,
		L.}, \bibinfo{author}{Wang, Y.}, \bibinfo{author}{Chen, J.},
	\bibinfo{year}{2020}.
	\newblock \bibinfo{title}{How distance transform maps boost segmentation cnns:
		An empirical study}, in: \bibinfo{editor}{Arbel, T.},
	\bibinfo{editor}{Ben~Ayed, I.}, \bibinfo{editor}{de~Bruijne, M.},
	\bibinfo{editor}{Descoteaux, M.}, \bibinfo{editor}{Lombaert, H.},
	\bibinfo{editor}{Pal, C.} (Eds.), \bibinfo{booktitle}{Proceedings of the
		Third Conference on Medical Imaging with Deep Learning},
	\bibinfo{publisher}{PMLR}. pp. \bibinfo{pages}{479--492}.
	\newblock \URLprefix \url{https://proceedings.mlr.press/v121/ma20b.html}.
	\bibitem[{Parzen(1962)}]{parzen62}
	\bibinfo{author}{Parzen, E.}, \bibinfo{year}{1962}.
	\newblock \bibinfo{title}{{On Estimation of a Probability Density Function and
			Mode}}.
	\newblock \bibinfo{journal}{The Annals of Mathematical Statistics}
	\bibinfo{volume}{33}, \bibinfo{pages}{1065 -- 1076}.
	\newblock \DOIprefix\doi{10.1214/aoms/1177704472}.
	\bibitem[{Pedrosa et~al.(2019)Pedrosa, Aresta, Ferreira, Rodrigues, Leitão,
		Carvalho, Rebelo, Negrão, Ramos, Cunha and Campilho}]{pedrosa19}
	\bibinfo{author}{Pedrosa, J.}, \bibinfo{author}{Aresta, G.},
	\bibinfo{author}{Ferreira, C.}, \bibinfo{author}{Rodrigues, M.},
	\bibinfo{author}{Leitão, P.}, \bibinfo{author}{Carvalho, A.S.},
	\bibinfo{author}{Rebelo, J.}, \bibinfo{author}{Negrão, E.},
	\bibinfo{author}{Ramos, I.}, \bibinfo{author}{Cunha, A.},
	\bibinfo{author}{Campilho, A.}, \bibinfo{year}{2019}.
	\newblock \bibinfo{title}{Lndb: A lung nodule database on computed tomography}.
	\newblock \URLprefix \url{https://arxiv.org/abs/1911.08434},
	\DOIprefix\doi{10.48550/ARXIV.1911.08434}.
	\bibitem[{Petersenn et~al.(2015)Petersenn, Richter, Broemel, Ritter,
		Deutschbein, Beil, Allolio, Fassnacht and Group}]{petersenn15}
	\bibinfo{author}{Petersenn, S.}, \bibinfo{author}{Richter, P.A.},
	\bibinfo{author}{Broemel, T.}, \bibinfo{author}{Ritter, C.O.},
	\bibinfo{author}{Deutschbein, T.}, \bibinfo{author}{Beil, F.U.},
	\bibinfo{author}{Allolio, B.}, \bibinfo{author}{Fassnacht, M.},
	\bibinfo{author}{Group, G.A.S.}, \bibinfo{year}{2015}.
	\newblock \bibinfo{title}{Computed tomography criteria for discrimination of
		adrenal adenomas and adrenocortical carcinomas: analysis of the german acc
		registry}.
	\newblock \bibinfo{journal}{European journal of endocrinology}
	\bibinfo{volume}{172}, \bibinfo{pages}{415--422}.
	\bibitem[{Phan et~al.(2019)Phan, Vo and Phan}]{phan18}
	\bibinfo{author}{Phan, A.C.}, \bibinfo{author}{Vo, V.Q.},
	\bibinfo{author}{Phan, T.C.}, \bibinfo{year}{2019}.
	\newblock \bibinfo{title}{A hounsfield value-based approach for automatic
		recognition of brain haemorrhage}.
	\newblock \bibinfo{journal}{Journal of Information and Telecommunication}
	\bibinfo{volume}{3}, \bibinfo{pages}{196--209}.
	\newblock \URLprefix \url{https://doi.org/10.1080/24751839.2018.1547951},
	\DOIprefix\doi{10.1080/24751839.2018.1547951}.
	\bibitem[{Ryu et~al.(2021)Ryu, Shin, Lee, Lee, Kang and Yi}]{ryu21}
	\bibinfo{author}{Ryu, H.}, \bibinfo{author}{Shin, S.Y.}, \bibinfo{author}{Lee,
		J.Y.}, \bibinfo{author}{Lee, K.M.}, \bibinfo{author}{Kang, H.j.},
	\bibinfo{author}{Yi, J.}, \bibinfo{year}{2021}.
	\newblock \bibinfo{title}{Joint segmentation and classification of hepatic
		lesions in ultrasound images using deep learning}.
	\newblock \bibinfo{journal}{European radiology} \bibinfo{volume}{31},
	\bibinfo{pages}{8733--8742}.
	\bibitem[{Shin et~al.(2015)Shin, Lee, Yun, Jung, Heo, Kim and Lee}]{shin15}
	\bibinfo{author}{Shin, S.Y.}, \bibinfo{author}{Lee, S.}, \bibinfo{author}{Yun,
		I.D.}, \bibinfo{author}{Jung, H.Y.}, \bibinfo{author}{Heo, Y.S.},
	\bibinfo{author}{Kim, S.M.}, \bibinfo{author}{Lee, K.M.},
	\bibinfo{year}{2015}.
	\newblock \bibinfo{title}{A novel cascade classifier for automatic
		microcalcification detection}.
	\newblock \bibinfo{journal}{PLOS ONE} \bibinfo{volume}{10},
	\bibinfo{pages}{1--22}.
	\newblock \URLprefix \url{https://doi.org/10.1371/journal.pone.0143725},
	\DOIprefix\doi{10.1371/journal.pone.0143725}.
	\bibitem[{Shin et~al.(2019)Shin, Lee, Yun, Kim and Lee}]{shin19}
	\bibinfo{author}{Shin, S.Y.}, \bibinfo{author}{Lee, S.}, \bibinfo{author}{Yun,
		I.D.}, \bibinfo{author}{Kim, S.M.}, \bibinfo{author}{Lee, K.M.},
	\bibinfo{year}{2019}.
	\newblock \bibinfo{title}{Joint weakly and semi-supervised deep learning for
		localization and classification of masses in breast ultrasound images}.
	\newblock \bibinfo{journal}{IEEE Transactions on Medical Imaging}
	\bibinfo{volume}{38}, \bibinfo{pages}{762--774}.
	\newblock \DOIprefix\doi{10.1109/TMI.2018.2872031}.
	\bibitem[{Shin et~al.(2023)Shin, Shen, Wank and Summers}]{shin23_spie}
	\bibinfo{author}{Shin, S.Y.}, \bibinfo{author}{Shen, T.C.},
	\bibinfo{author}{Wank, S.A.}, \bibinfo{author}{Summers, R.M.},
	\bibinfo{year}{2023}.
	\newblock \bibinfo{title}{{Improving small lesion segmentation in CT scans
			using intensity distribution supervision: application to small bowel
			carcinoid tumor}}, in: \bibinfo{editor}{Iftekharuddin, K.M.},
	\bibinfo{editor}{Chen, W.} (Eds.), \bibinfo{booktitle}{Medical Imaging 2023:
		Computer-Aided Diagnosis}, \bibinfo{organization}{International Society for
		Optics and Photonics}. \bibinfo{publisher}{SPIE}. p.
	\bibinfo{pages}{124651S}.
	\newblock \URLprefix \url{https://doi.org/10.1117/12.2651979},
	\DOIprefix\doi{10.1117/12.2651979}.
	\bibitem[{Sudre et~al.(2017)Sudre, Li, Vercauteren, Ourselin and
		Jorge~Cardoso}]{sudre17}
	\bibinfo{author}{Sudre, C.H.}, \bibinfo{author}{Li, W.},
	\bibinfo{author}{Vercauteren, T.}, \bibinfo{author}{Ourselin, S.},
	\bibinfo{author}{Jorge~Cardoso, M.}, \bibinfo{year}{2017}.
	\newblock \bibinfo{title}{Generalised dice overlap as a deep learning loss
		function for highly unbalanced segmentations}, in: \bibinfo{editor}{Cardoso,
		M.J.}, \bibinfo{editor}{Arbel, T.}, \bibinfo{editor}{Carneiro, G.},
	\bibinfo{editor}{Syeda-Mahmood, T.}, \bibinfo{editor}{Tavares, J.M.R.},
	\bibinfo{editor}{Moradi, M.}, \bibinfo{editor}{Bradley, A.},
	\bibinfo{editor}{Greenspan, H.}, \bibinfo{editor}{Papa, J.P.},
	\bibinfo{editor}{Madabhushi, A.}, \bibinfo{editor}{Nascimento, J.C.},
	\bibinfo{editor}{Cardoso, J.S.}, \bibinfo{editor}{Belagiannis, V.},
	\bibinfo{editor}{Lu, Z.} (Eds.), \bibinfo{booktitle}{Deep Learning in Medical
		Image Analysis and Multimodal Learning for Clinical Decision Support},
	\bibinfo{publisher}{Springer International Publishing},
	\bibinfo{address}{Cham}. pp. \bibinfo{pages}{240--248}.
	\bibitem[{Summers et~al.(2006)Summers, Huang, Yao, Campbell, Dempsey, Dwyer,
		Franaszek, Brickman, Bitter, Petrick and Hara}]{summers06}
	\bibinfo{author}{Summers, R.M.}, \bibinfo{author}{Huang, A.},
	\bibinfo{author}{Yao, J.}, \bibinfo{author}{Campbell, S.R.},
	\bibinfo{author}{Dempsey, J.E.}, \bibinfo{author}{Dwyer, A.J.},
	\bibinfo{author}{Franaszek, M.}, \bibinfo{author}{Brickman, D.S.},
	\bibinfo{author}{Bitter, I.}, \bibinfo{author}{Petrick, N.},
	\bibinfo{author}{Hara, A.K.}, \bibinfo{year}{2006}.
	\newblock \bibinfo{title}{Assessment of polyp and mass histopathology by
		intravenous contrast–enhanced ct colonography}.
	\newblock \bibinfo{journal}{Academic Radiology} \bibinfo{volume}{13},
	\bibinfo{pages}{1490--1495}.
	\newblock \DOIprefix\doi{https://doi.org/10.1016/j.acra.2006.09.051}.
	\bibitem[{Tang et~al.(2020)Tang, Tang, Zhu, Xiao and Summers}]{tang20}
	\bibinfo{author}{Tang, Y.}, \bibinfo{author}{Tang, Y.}, \bibinfo{author}{Zhu,
		Y.}, \bibinfo{author}{Xiao, J.}, \bibinfo{author}{Summers, R.M.},
	\bibinfo{year}{2020}.
	\newblock \bibinfo{title}{E$^2${N}et: An edge enhanced network for accurate
		liver and tumor segmentation on ct scans}, in: \bibinfo{editor}{Martel,
		A.L.}, \bibinfo{editor}{Abolmaesumi, P.}, \bibinfo{editor}{Stoyanov, D.},
	\bibinfo{editor}{Mateus, D.}, \bibinfo{editor}{Zuluaga, M.A.},
	\bibinfo{editor}{Zhou, S.K.}, \bibinfo{editor}{Racoceanu, D.},
	\bibinfo{editor}{Joskowicz, L.} (Eds.), \bibinfo{booktitle}{Medical Image
		Computing and Computer Assisted Intervention -- MICCAI 2020},
	\bibinfo{publisher}{Springer International Publishing},
	\bibinfo{address}{Cham}. pp. \bibinfo{pages}{512--522}.
	\bibitem[{Weninger et~al.(2020)Weninger, Liu and Merhof}]{weninger20}
	\bibinfo{author}{Weninger, L.}, \bibinfo{author}{Liu, Q.},
	\bibinfo{author}{Merhof, D.}, \bibinfo{year}{2020}.
	\newblock \bibinfo{title}{Multi-task learning for brain tumor segmentation},
	in: \bibinfo{editor}{Crimi, A.}, \bibinfo{editor}{Bakas, S.} (Eds.),
	\bibinfo{booktitle}{Brainlesion: Glioma, Multiple Sclerosis, Stroke and
		Traumatic Brain Injuries}, \bibinfo{publisher}{Springer International
		Publishing}, \bibinfo{address}{Cham}. pp. \bibinfo{pages}{327--337}.
	\bibitem[{Zhao et~al.(2021)Zhao, Chen and Wang}]{zhao21}
	\bibinfo{author}{Zhao, Z.}, \bibinfo{author}{Chen, H.}, \bibinfo{author}{Wang,
		L.}, \bibinfo{year}{2021}.
	\newblock \bibinfo{title}{A coarse-to-fine framework for the 2021 kidney and
		kidney tumor segmentation challenge}.
	\newblock \URLprefix \url{https://openreview.net/forum?id=6Py5BNBKoJt}.
	
\end{thebibliography}






\end{document}